%% file: a2255_udf.tex
\documentclass[longauth]{aa}
\input{latex_mycommands.txt}

\usepackage{graphicx}
\usepackage{amssymb}
\usepackage{multirow,bigdelim}
\usepackage{txfonts}
\usepackage[export]{adjustbox}
\usepackage{hyperref}
\hypersetup{colorlinks,citecolor=blue,filecolor=black,linkcolor=red,urlcolor=black}
\begin{document} 

\title{The topology of the magnetic field in Abell 2255 out to its virial radius}
\subtitle{Results from the LOFAR Galaxy Cluster Ultra-Deep Field}

\authorrunning{A. Botteon et al.} 
\titlerunning{Magnetic fields in the LOFAR Galaxy Cluster Ultra-Deep Field}

\author{A. Botteon\inst{\ref{ira}},
R. J. van Weeren\inst{\ref{leiden}},
Y. Hu\inst{\ref{ias}},
F. Vazza\inst{\ref{unibo},\ref{ira}},
G. Brunetti\inst{\ref{ira}},
K. Rajpurohit\inst{\ref{cfa}},
A. Lazarian\inst{\ref{madison}},
T. W. Shimwell\inst{\ref{astron},\ref{leiden}},
E. De Rubeis\inst{\ref{hamburg},\ref{ira}},
M. Balboni\inst{\ref{unibo},\ref{iasf}},
A. Bonafede\inst{\ref{unibo},\ref{ira}},
R. Cassano\inst{\ref{ira}},
G. Di Gennaro\inst{\ref{ira}},
F. Gastaldello\inst{\ref{iasf}},
M. J. Hardcastle\inst{\ref{herts}},
A. Ignesti\inst{\ref{prague}},
and
H. J. A. R\"{o}ttgering\inst{\ref{leiden}}
}

\institute{
INAF - IRA, via P.~Gobetti 101, 40129 Bologna, Italy \label{ira} \\
\email{andrea.botteon@inaf.it} 
\and
Leiden Observatory, Leiden University, PO Box 9513, 2300 RA Leiden, The Netherlands \label{leiden}
\and
Institute for Advanced Study, 1 Einstein Drive, Princeton, NJ 08540, USA \label{ias}
\and
Dipartimento di Fisica e Astronomia, Universit\`{a} di Bologna, via P.~Gobetti 93/2, 40129 Bologna, Italy \label{unibo}
\and
Center for Astrophysics Harvard \& Smithsonian, 60 Garden Street, Cambridge, MA 02138, USA \label{cfa}
\and
Department of Astronomy, University of Wisconsin-Madison, Madison, WI 53706, USA \label{madison}
\and
ASTRON, the Netherlands Institute for Radio Astronomy, Postbus 2, 7990 AA Dwingeloo, The Netherlands \label{astron}
\and
Hamburger Sternwarte, Universit\"{a}t Hamburg, Gojenbergsweg 112, 21029 Hamburg, Germany \label{hamburg}
\and
INAF - IASF Milano, via A.~Corti 12, 20133 Milano, Italy \label{iasf}%
\and
Department of Physics, Astronomy and Mathematics, University of Hertfordshire, College Lane, Hatfield AL10 9AB, UK  \label{herts}
\and
Astronomical Institute of the Czech Academy of Sciences, Bo\v{c}n\'i II 1401, 14100, Prague, Czech Republic \label{prague}
}

\date{Received XXX; accepted YYY}

\abstract
{We present the \lofar\ Galaxy Cluster Ultra-Deep Field, in which 336~h of \lofar\ observations at 120--168 MHz have been collected on the nearby ($z=0.080$) cluster Abell 2255. This massive and merging system is known to host spectacular radio emission from both cluster galaxies and the intracluster medium. Previous \lofar\ observations revealed pervasive diffuse synchrotron emission extending from the cluster center to its dynamically active outskirts, tracing relativistic electrons propagating in large-scale magnetic fields. In this work, we present a set of new ultra-deep images at the central frequency of 144 MHz based on the 224~h of data with the best quality, which reach a sensitivity of 24 \mujyb\ at $7.1\arcsec \times 4.3\arcsec$ resolution. These images represent the deepest radio observations of a galaxy cluster obtained to date and provide a glimpse of what should be routinely observed in clusters with \ska-Low in the near future. Using these data, we investigate the topology of the cluster magnetic field out to its virial radius by applying the synchrotron intensity gradient technique. We find that the inferred magnetic field exhibits preferential orientations in distinct regions of the cluster, such as in the radio halo extensions (bridges) and in the relics, suggesting that the dynamics of the cluster formation process is shaping the large-scale magnetic field. This interpretation is supported by the comparison with the magnetic field orientation obtained from cosmological magnetohydrodynamic simulations. This work provides the first indication of a coherent, large-scale magnetic field topology across an entire galaxy cluster, from core to outskirts, and demonstrates the unique power of ultra-deep, low-frequency observations to trace the structure of cluster magnetic fields on megaparsec scales, thereby probing the magnetization of the large-scale structure of the Universe.
}

\keywords{acceleration of particles -- magnetic fields -- radiation mechanisms: non-thermal -- galaxies: clusters: intracluster medium -- galaxies: clusters: general -- galaxies: clusters: individual: A2255}

\maketitle

\section{Introduction}

Magnetic fields are ubiquitous in the Universe \citep[\eg][]{brandenburg05rev, widrow12rev}. In galaxy clusters, they are thought to permeate the entire intracluster medium (ICM): a hot, diffuse, and rarefied plasma that fills the cluster volume and emits thermal bremsstrahlung radiation in the X-ray band \citep[\eg][]{sarazin86rev, carilli02rev, govoni04rev}. In combination with cosmic ray electrons, magnetic fields in the ICM give rise to cluster-scale nonthermal synchrotron emission at radio wavelengths \citep[\eg][]{vanweeren19rev}. In particular, radio relics and halos, which are predominately observed in dynamically disturbed clusters, demonstrate that a fraction of the kinetic energy dissipated by shocks and turbulence during the hierarchical assembly of galaxy clusters is channeled into nonthermal components on megaparsec scales \citep[\eg][]{brunetti14rev}. Despite the fundamental role that magnetic fields play in governing key aspects of the ICM (micro)physics, their origin, evolution, strength, and topology are still poorly constrained due to observational limitations. This knowledge gap limits our understanding of galaxy cluster evolution, the dynamics and thermalization of the ICM, the transport and acceleration of cosmic rays, and the generation of nonthermal emission in the form of radio halos and relics. \\
\indent
Magnetic fields in the ICM are primarily probed through Faraday rotation measure (RM) studies of polarized background or embedded radio sources, which are sensitive to the line-of-sight component of the magnetic field weighted by the thermal electron density \citep[\eg][]{clarke04faraday}. Targeted observations of nearby clusters and statistical studies of large samples suggest that magnetic fields in the ICM typically (i) reach strengths of a few \muG\ in cluster cores, (ii) decline toward the outskirts, and (iii) are correlated with the thermal electron density \citep[\eg][]{feretti99a119, murgia04, bonafede10, bonafede13, vacca12, osinga25}. Most of these studies were carried out with the \vla\ and were limited only to a handful of polarized sources per cluster. However, new-generation instruments operating at frequencies $\sim$1 GHz, such as \meerkat\ and \askap, are now able to recover tens of RMs per square degree, significantly increasing the density of RM grids to probe cluster magnetic fields \citep[\eg][]{anderson21, loi25, pagliotta25, alonsolopez26, khadir26}. While this represents a major step forward, the application of Faraday rotation still faces intrinsic limitations and degeneracies \citep[\eg][]{johnson20} and, at low frequencies ($\lesssim$200 MHz), it becomes ineffective due to Faraday depolarization effects \citep{burn66}, which severely hinders the detection of polarized sources behind and within clusters \citep[\eg][]{farnsworth11}. An alternative approach to infer the magnetic field properties (in particular, the orientation) in astrophysical sources, known as synchrotron intensity gradient (SIG) technique, has been proposed \citep{lazarian17} and recently applied to galaxy clusters \citep{hu24clusters, hu25statistics}. This method is based on the magnetohydrodynamic (MHD) turbulence theory and is independent of Faraday rotation. The application of SIG to cluster diffuse sources in head-on cluster mergers suggests that the large-scale ICM magnetic field is preferentially aligned along the merger axis, providing, for the first time, a way to probe the large-scale structure and dynamics of the cluster field \citep{hu24clusters}. \\
\indent
Thanks to recent advances in radio astronomy, driven in particular by the development of pathfinders and precursors of the \skaE\ (\ska), the fidelity and sensitivity of radio images have dramatically improved in recent years. Major progress has been achieved at low radio frequencies with \lofar, which is currently the most powerful radio telescope operating at $\lesssim$200~MHz. The deepest \lofar\ observations at 120--168 MHz have targeted four well-studied extragalactic fields [Bo\"otes, Lockman Hole, ELAIS-N1 (European Large Area ISO Survey-North 1), and NEP (North Ecliptic Pole)] which are characterized by extensive multiwavelength coverage and were primarily designed to investigate the formation and evolution of galaxies \citep{best23}. Among these, ELAIS-N1 is the deepest field, with a total integration time of 505~h and a noise level of 10.7~\mujyb\ at $6\arcsec$ resolution; the final data products have been released recently \citep{shimwell25}. Highly sensitive radio images are also essential for the study of cosmic ray electrons and magnetic fields in galaxy clusters as the synchrotron emission originating from these nonthermal components is diffuse and characterized by a very low surface brightness. Taking advantage of the multibeam capability of \lofar, the NEP field and the field containing the nearby galaxy cluster Abell 2255 (hereafter, A2255) were co-observed. As a result, A2255 effectively became the \lofar\ Galaxy Cluster Deep Field. First results, based on 72~h of observations, were presented in \citet{botteon22a2255}. Over the past years, the total integration time on the cluster field has increased to 336~h, forming the Ultra-Deep Field dataset that we present in this paper. \\
\indent
A2255 is a galaxy cluster at $z = 0.080$ \citep{golovich19atlas} that is well known for its rich radio phenomenology, arising from both cluster galaxies and the ICM, whose emission has been extensively studied in the past with \wsrt\ and \vla\ \citep{jaffe79, harris80eight, miller03a2255, feretti97a2255, pizzo08, pizzo09, pizzo11, govoni05, govoni06}. The observed emission includes tailed radio galaxies associated with active galactic nuclei (AGN), as well as diffuse emission in the form of a radio halo and multiple relics generated during the ongoing merger activity of the cluster, which is highlighted also by observations in optical and X-ray bands \citep[\eg][]{burns95, davis98, sakelliou06, shim11, akamatsu17a2255, golovich19atlas}. Interest on the radio emission from A2255 was renewed when observations from the \lotssE\ \citep[\lotss;][]{shimwell19, shimwell22, shimwell26} covered its field. Leveraging the survey data, which had only 8~h of integration time, \citet{botteon20a2255} revealed the remarkable complexity of the radio sources in A2255, simultaneously recovering both small-scale structures (such as filaments within radio galaxies and the halo) and large-scale diffuse emission. With deeper \lofar\ observations, 72~h at 120--168 MHz and 72~h at 22--70 MHz, the analysis focused mainly on the most extended diffuse radio emission, which was found to expand from the cluster center up to its virial radius, demonstrating the presence of nonthermal components distributed over a largest projected linear size of $\sim$5 Mpc \citep{botteon22a2255}. The 72~h observation at 120--168 MHz (\ie\ the anticipated Deep Field data) were also used to study the nonthermal radio emission in ram-pressure stripped galaxies in the system \citep{ignesti23a2255} and, by including the international \lofar\ baselines in the analysis (previously neglected), to investigate cluster radio galaxies and the network of filaments down to sub-arcsecond resolution \citep{derubeis25a2255, derubeis26}. \\
\indent
With the total 336~h integration of the Ultra-Deep Field, we can obtain highly sensitive low-frequency radio images of A2255, which are crucial for recovering the most extended and faint diffuse emission from the system at high significance. By taking advantage of the pervasive large-scale synchrotron emission in the cluster, we apply the SIG technique to investigate the topology of the ICM magnetic field out to the cluster virial radius. The combination of this unique dataset and this novel approach currently provides the only viable way to probe the magnetic field structure from the center to the outskirts of a cluster. \\
\indent
This paper is structured as follows. In Section~\ref{sec:reduction}, we describe the data and processing of the observations constituting the \lofar\ Galaxy Cluster Ultra-Deep Field. In Section~\ref{sec:udf}, we present the new images of A2255. In Section~\ref{sec:sig}, we introduce the SIG and apply the method on the \lofar\ images of A2255. In Sections~\ref{sec:discussion} and \ref{sec:conclusions} we discuss and summarize the results of our analysis. \\
\indent
Hereafter, we adopt a $\Lambda$ cold dark matter cosmology with $\omegal = 0.7$, $\omegam = 0.3$, and $\hzero = 70$ \kmsmpc, in which 1\arcsec\ corresponds to 1.512 kpc at the cluster redshift.%

\section{Data reduction}\label{sec:reduction}

\subsection{Observation setup}

The \lofar\ Galaxy Cluster Ultra-Deep Field dataset is composed of observations from two projects: LC12\_027 (PI: R. van   Weeren) and LT16\_005 (PI: P. Best). The final on-source integration time of 336~h was obtained by combining 42 individual observations carried out with identical observational setup and duration. Each observation consisted of an 8~h integration on the target, bracketed by two 10~min scans on the flux density calibrators 3C295 and 3C48, and was performed in the HBA\_DUAL\_INNER mode over the frequency range 120--168 MHz. In these observations, one \lofar\ beam was pointed at A2255, while the second beam simultaneously covered the NEP field. The first 72~h of observations were conducted in 2019 under project LC12\_027 and constitute the Deep Field dataset already published (in \citealt{botteon22a2255} for A2255, and in \citealt{bondi24} for NEP). The remaining 264~h of observations were obtained between 2022 and 2023 under project LT16\_005. While international stations were present in all observations, in this work we focus only on the data recorded with the Dutch array. The total size of data retrieved from the \lofar\ Long Term Archive (LTA) for this project amounts to $\sim$200 TB. The complete list of the observations is provided in Tab.~\ref{tab:radio_obs}.  \\
\indent
The data processing closely followed the procedures described by \citet{tasse21} and \citet{shimwell19, shimwell22, shimwell26} for the \lotss\ Deep Fields and the wide-area survey; the same approach was also adopted by \citet{botteon22a2255} in the analysis of the initial 72~h observations of A2255. In the following sections, we briefly outline the main processing steps.

\subsection{Direction-independent calibration (PreFactor)}

Data were retrieved from the LTA, where they underwent a first round of flagging using \aoflagger\ \citep{offringa12}, averaging with \dpppE\ \citep[\texttt{DP3};][]{vandiepen18}, and compression with \dysco\ \citep{offringa16}. Following the download, they were corrected for direction-independent instrumental and systematic effects. This initial step was performed using the \texttt{PreFactor}\footnote{\url{https://github.com/lofar-astron/prefactor}} calibrator and target pipelines\footnote{Now known as \lofar\ Initial Calibration (\texttt{LINC}) pipelines.} \citep{vanweeren16calibration, williams16, degasperin19}. In the \texttt{PreFactor} calibrator pipeline, data are averaged and low-quality data are flagged. The calibrator data are then corrected for clock offsets between different stations, ionospheric Faraday rotation, phase offset between the XX and YY correlations (polarization alignment), and amplitude gains. The derived calibration solutions are subsequently applied to the target field using the \texttt{PreFactor} target pipeline. At this stage, phase calibration is performed against a sky model from the \tgssE\ \citep[\tgss;][]{intema17}, together with the additional flagging of bad data and the removal of the international stations (not used in our analysis). \\
\indent
For the observations carried out under project LC12\_027, we adopted the calibrator solutions derived on 3C48 by \citet{botteon22a2255}. However, for two observations affected by poorer data quality, the \texttt{PreFactor} calibrator pipeline was rerun on 3C295, and the corresponding solutions were used. In all cases, the selected calibrator solutions were transferred to the target field by rerunning the \texttt{PreFactor} target pipeline on the target data. The processing of the LT16\_005 observations initially started using 3C48 as the primary calibrator. However, after identifying generally noisy and incoherent solutions, we decided to systematically run the \texttt{PreFactor} calibrator pipeline on both calibrators and ultimately decided to adopt 3C295 for all observations. During this step, we labeled four observations as failed (see Tab.~\ref{tab:radio_obs}) because the \texttt{PreFactor} calibrator pipeline rejected about half of the data due to poor quality; these observations were therefore discarded. All four failed observations were obtained in July 2022. At a later stage of the analysis, we found that two of the remaining three observations from the same month also had to be discarded due to insufficient data quality. \\
\indent
This step of the analysis was carried out on the HOTCAT \citep{bertocco20, taffoni20} High Performance Computing (HPC) cluster at INAF--OATs, which hosts 4 large computing nodes\footnote{Each equipped with four Intel(R) Xeon(R) Gold 5118 processors and 512~GB of RAM.} dedicated to \lofar\ data analysis.

\subsection{Direction-dependent calibration (ddf-pipeline)}

A first round of direction-dependent calibration was performed using \texttt{ddf-pipeline}\footnote{\url{https://github.com/mhardcastle/ddf-pipeline}}, which was run independently on the 38 observations that successfully completed the \texttt{PreFactor} stage, adopting a common input sky model and facet layout (see below). This pipeline relies on \texttt{killMS} \citep{tasse14arx, smirnov15} to solve for direction-dependent ionospheric effects, and on \texttt{DDFacet} \citep{tasse18} to apply the corresponding calibration solutions during imaging. Leveraging prior knowledge of the quality of the first 72~h of observations \citep{botteon22a2255, derubeis25a2255, derubeis26}, the best-quality dataset among the first 9 observing nights (\ie\ SAS ID: 747611) was selected as the reference input sky model and facet layout for the remaining observations to speed up the processing \citep[see][]{tasse21, shimwell25}. The final full field-of-view (FoV) images produced by the pipeline at 6\arcsec-resolution, corrected for both direction-dependent and direction-independent effects, were then used for a first assessment of the quality of the individual observations. \\
\indent
This step of the analysis was carried out on the \lofar-Pleiadi HPC system at INAF--IRA, which comprises of 36 computing nodes\footnote{Each equipped with two Intel(R) Xeon(R) E5-2697 v4 processors and 256~GB of RAM.}.

\subsection{Refinement of the calibration (facetselfcal)}

\begin{figure*}
  \centering
  \includegraphics[width=\hsize,trim={0cm 0cm 0cm 0cm},clip,valign=c]{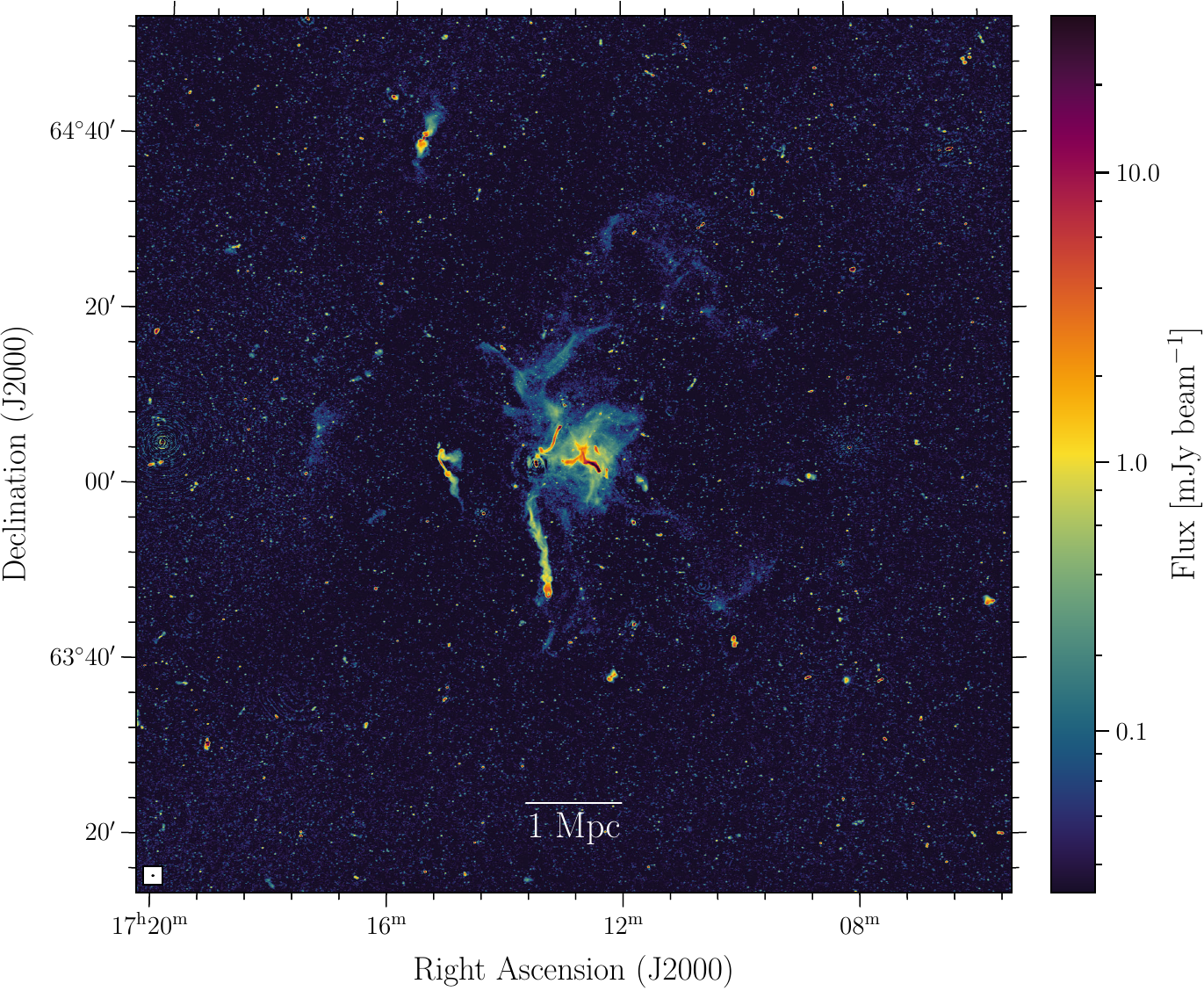}
  \caption{Wide-field image of A2255 at $7.1\arcsec\ \times 4.3\arcsec$ resolution. The noise is $\sigma_{\rm rms} = 24$ \mujyb\ and the color scale has a logarithmic stretch from 0.5 to 1500$\sigma_{\rm rms}$. The radio beam is shown in the bottom left corner.}
  \label{fig:robust-0.5}
\end{figure*}

\begin{figure*}
  \centering
  \includegraphics[width=\hsize,trim={0cm 0cm 0cm 0cm},clip,valign=c]{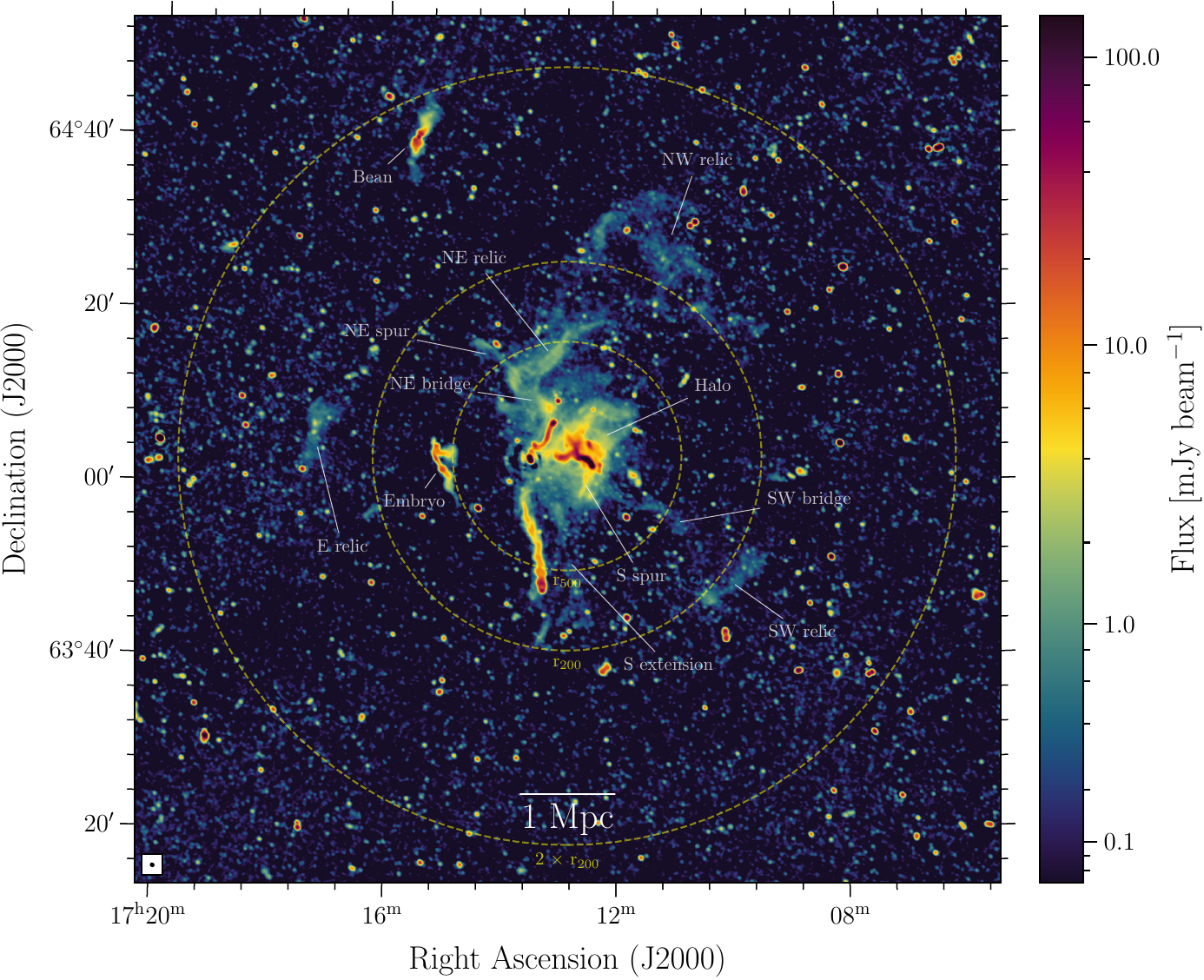}
  \caption{Wide-field low-resolution image of A2255 obtained with a 15\arcsec-Gaussian taper. The image resolution and noise are $20.7\arcsec\ \times 18.0\arcsec$ and $\sigma_{\rm rms} = 93$ \mujyb, respectively. The color scale has a logarithmic stretch from 0.5 to 1500$\sigma_{\rm rms}$. The radio beam is shown in the bottom left corner. Circles denote different characteristic radii. The main radio sources in the field are labeled following previous literature studies, with the exception of the candidate E relic which was not reported earlier.}
  \label{fig:taper15}
\end{figure*}

\begin{figure*}
  \centering
  \includegraphics[width=\hsize,trim={0cm 0cm 0cm 0cm},clip,valign=c]{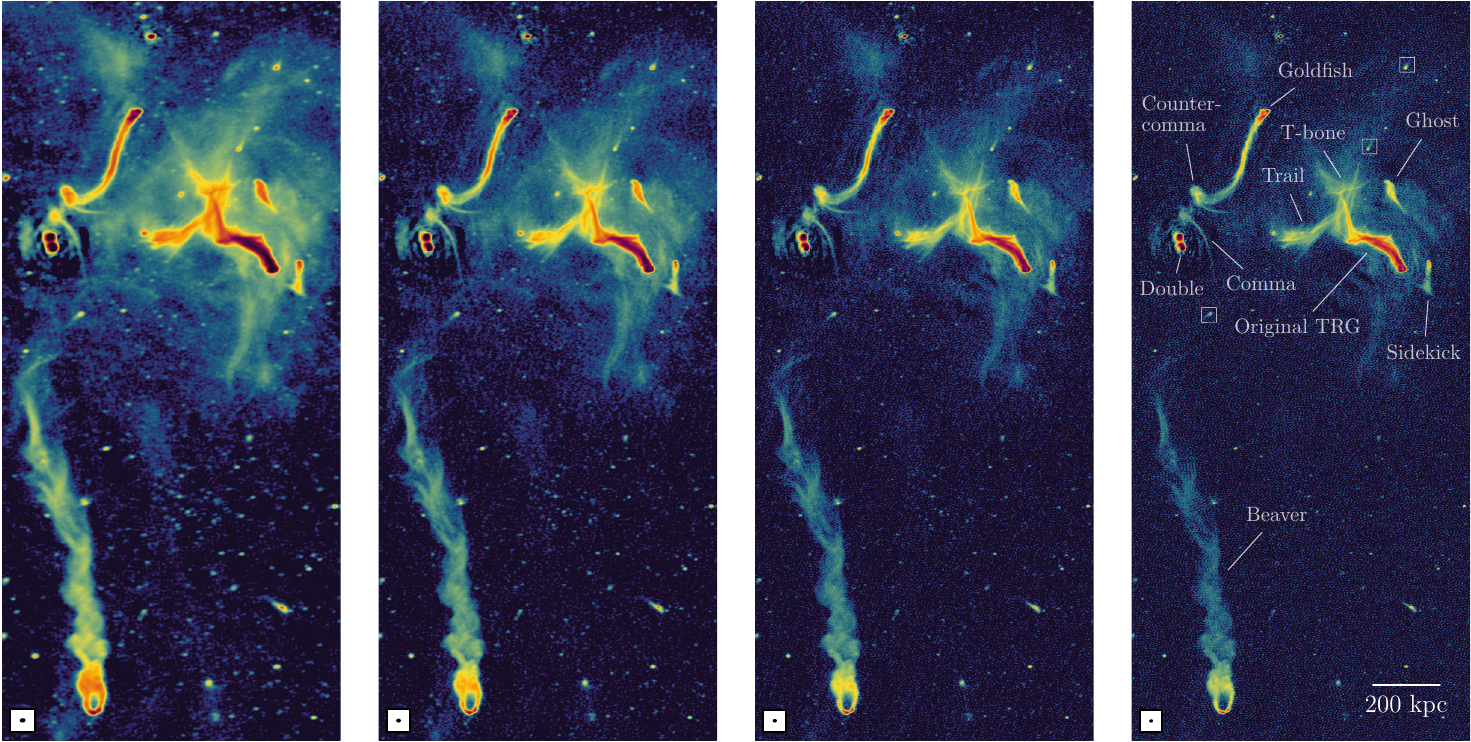}
  \caption{Zoom-in of the complex central region of A2255 at progressively higher resolution. From \textit{left} to \textit{right}, images were obtained with robust weighting of the visibilities of $-$0.5, $-$1.0, $-$1.5, and $-$2.0. The corresponding resolutions ($\sigma_{\rm rms}$ values) are $7.1\arcsec\ \times 4.3\arcsec$ (24 \mujyb), $4.9\arcsec\ \times 3.5\arcsec$ (25 \mujyb), $4.0\arcsec\ \times 2.8\arcsec$ (42 \mujyb), and $3.3\arcsec\ \times 2.2\arcsec$ (72 \mujyb). The color scale has a common logarithmic stretch from 0.5 to 2000 $\times$ 24 \mujyb. The radio beams are shown in the bottom left corners. The main radio sources in this region are labeled following previous literature studies. Squared regions denote three ram-pressure stripped star-forming galaxies.}
  \label{fig:robust_gallery}
\end{figure*}

We found that approximately 20--25\% of the observations processed with \texttt{ddf-pipeline} still suffered from strong ionospheric distortions, resulting in very poor image quality. To further improve the calibration in the central region of the field, we therefore employed the ``extraction and recalibration'' scheme introduced by \citet{vanweeren21}. Here, the FoV used for calibration (and thus for the subsequent analysis) is restricted to a smaller region compared to that calibrated by \texttt{ddf-pipeline} (\ie\ $8.96\deg \times 8.96\deg$). Specifically, hereafter we focused on a $2.0\deg \times 2.0\deg$ field centered on A2255. All sources outside this region were subtracted from the visibilities using the sky model provided by the \texttt{ddf-pipeline} (this is the so-called ``extraction'' step). %
The resulting $2.0\deg \times 2.0\deg$ FoV encompasses a region extending to roughly three times the cluster virial radius and is four times larger than the 1 deg$^2$ field analyzed in \citet{botteon22a2255}. A first run of \texttt{facetselfcal}\footnote{\url{https://github.com/rvweeren/lofar_facet_selfcal}} (which performs the ``recalibration'' step) was performed independently on each dataset to validate the data quality inferred from the \texttt{ddf-pipeline} images. We confirmed that 10 observations had to be rejected due to poor data quality, 8 were classified as having medium quality, and the 20 remaining datasets were labeled as good quality (see Tab.~\ref{tab:radio_obs}). This individual extraction and recalibration of the 38 datasets was also carried out on the \lofar-Pleiadi HPC system at INAF--IRA. \\
\indent
The final calibration with \texttt{facetselfcal} was then performed jointly on the 28 datasets classified as either good or medium quality on a dedicated large \lofar\ node\footnote{Equipped with two AMD EPYC 7413 processors and 1~TB of RAM.} at INAF--IRA. As a result, the final analysis is based on a total of 224~h of observations, implying that 112~h of data from the original Ultra-Deep Field dataset (\ie\ 33\% of the total) were discarded due to inferior data quality. We attribute the reduced quality of these rejected observations primarily to increased solar activity, which strongly affects low-frequency radio observations and was at minimum in 2019, when the first 72~h of data were collected. \\
\indent
While one of the main reasons for adopting the extraction step is to enable recalibration on a smaller FoV, where direction-independent calibration remains effective \citep[\eg][]{vanweeren21, botteon22dr2}, for larger extracted fields, such as that used for A2255, corrections for direction-dependent effects are still required. We therefore performed a new, refined, direction-dependent calibration with \texttt{facetselfcal} using 14 facets, for which the facet solutions were derived following the method described in \citet{dejong22}. Compared to the initial \texttt{ddf-pipeline} processing, which employs 45 facets over a FoV of $8.96\deg \times 8.96\deg$, adopting 14 facets over the $2.0\deg \times 2.0\deg$ extracted field of A2255 enables a more precise calibration of the ionosphere, as the solutions are optimized over smaller regions of the sky, allowing a better correction of local phase variations. While \texttt{ddf-pipeline} relies on \texttt{killMS} and \texttt{DDFacet}, \texttt{facetselfcal} instead uses \texttt{DP3} for calibration and \wsclean\ \citep{offringa14} for imaging. An initial sky model was obtained from a full calibration run on a single observation, using the best-quality dataset (\ie\ SAS ID: 747611), and was adopted as starting model for the joint calibration of the final dataset to speed up the processing.

\subsection{Final imaging}

Final imaging was performed with \wsclean\ by jointly deconvolving the 28 datasets by applying the direction-dependent calibration solutions obtained with \texttt{facetselfcal} during imaging. During imaging, we employed cleaning masks to guide the deconvolution and adopted the \texttt{wgridder} algorithm \citep{arras21, ye22} together with the multiscale multi-frequency deconvolution option \citep{offringa17}. Our reference image was obtained using \citet{briggs95} weighting with a robust parameter of $-0.5$ and an inner \uv\ cut of 60$\lambda$ (equivalent to an angular scale of 57.30\arcmin). Lower-resolution images were produced by applying Gaussian tapers to the visibilities, keeping the robust parameter at $-0.5$, while higher-resolution images were obtained using more negative robust weightings. To assess the impact of short baselines on the recovery of extended diffuse emission, we also produced images using inner \uv\ cuts ranging from 30$\lambda$ to 70$\lambda$ (\ie\ from 114.59\arcmin\ to 49.11\arcmin), in steps of 10$\lambda$ (see Appendix~\ref{app:images}). \\
\indent
In order to enhance the cluster diffuse radio emission, discrete sources were subtracted from the visibilities. This was achieved by filtering out large-scale emission on selected angular scales through the application of inner \uv\ cuts when producing the model images for subtraction. Our reference subtraction model was derived from an image obtained with an inner \uv\ cut of 5000$\lambda$ (equivalent to an angular scale of 0.69\arcmin), while additional models derived from images obtained with cuts of 1500, 2500, 8000, and 10000$\lambda$ (\ie\ from 2.29\arcmin\ to 0.34\arcmin) were produced to assess the robustness and quality of the subtraction (see Appendix~\ref{app:images}). All model images were visually inspected to remove possible cluster diffuse emission that may have been included in the models despite the applied \uv\ cuts. Thus, models were predicted into the visibilities using the direction-dependent solutions and subtracted from the data, yielding residual visibilities that contain only the unsubtracted components. These residual visibilities were subsequently imaged to produce the images with the discrete sources subtracted reported in the paper. As in \citet{botteon22a2255}, we did not attempt to subtract the extended and bright tailed radio galaxies in the cluster, due to the difficulty of reliably modeling and subtracting their emission in the \uv\ plane \citep[see also][]{botteon24a754, vanweeren24, rajpurohit25profiles}. These sources were instead masked in the subsequent analysis of the diffuse cluster emission. \\
\indent
All images have been corrected for the primary beam response and are reported at the central frequency of 144~MHz. The flux density scale has been aligned to that of \lotss-DR3 \citep{shimwell26}. %

\section{LOFAR Ultra-Deep Field images of A2255}\label{sec:udf}

\begin{figure*}
  \centering
  \includegraphics[width=\hsize,trim={0cm 0cm 0cm 0cm},clip,valign=c]{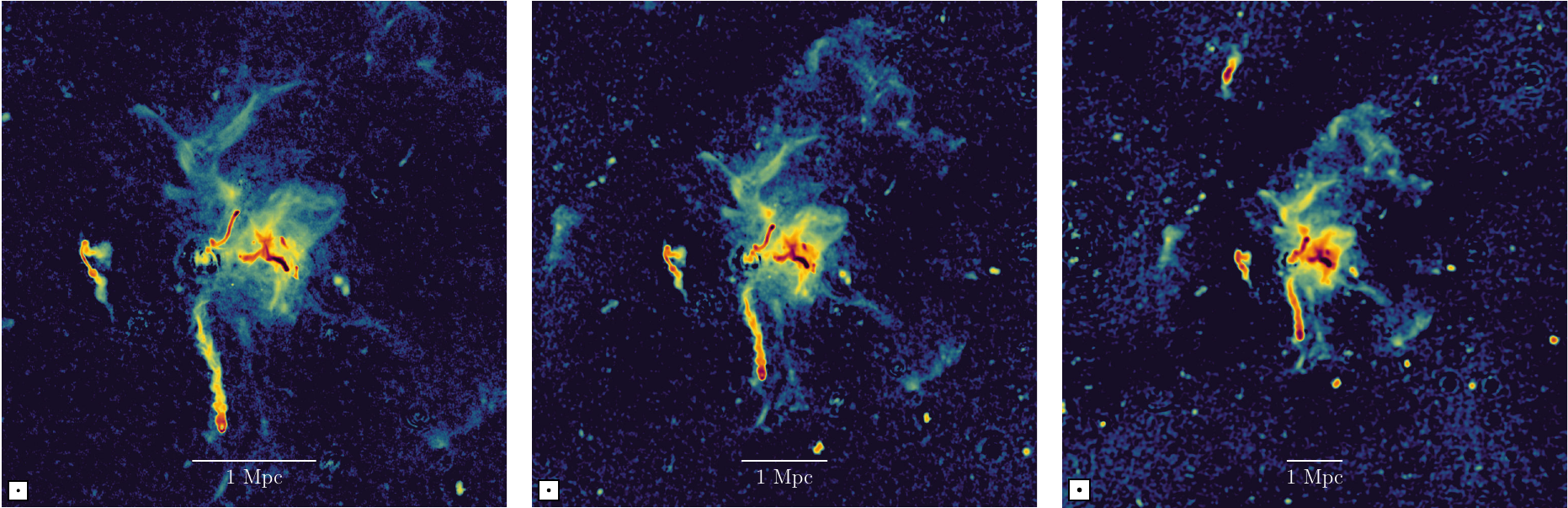}
  \caption{Images of A2255 with discrete sources subtracted. From \textit{left} to \textit{right}, images were smoothed to circular beams of 8\arcsec, 16\arcsec, and 35\arcsec, achieving noise values of $\sigma_{\rm rms} = 30$, 59, and 127 \mujyb. The displayed FoVs are 45, 65, and 100 arcmin$^2$. The color scale has a logarithmic stretch from 0.5 to 1500$\sigma_{\rm rms}$. The radio beams are shown in the bottom left corners.}
  \label{fig:sub_gallery}
\end{figure*}

In Figs.~\ref{fig:robust-0.5} and \ref{fig:taper15}, we show the 224~h \lofar\ images at 144 MHz of A2255 obtained with a robust weighting of $-$0.5 at full resolution and with a 15\arcsec-Gaussian taper applied, respectively. The first image has a noise of $\sigma_{\rm rms} = 24$ \mujyb\ and a resolution of $7.1\arcsec\ \times 4.3\arcsec$, while the other has a noise of $\sigma_{\rm rms} = 99$ \mujyb\ and a resolution of $20.7\arcsec\ \times 18.0\arcsec$. The two images cover a FoV of 100\arcmin\ $\times$ 100\arcmin, approximately equivalent to $2\times\rtwo$ or $3.4\times\rfive$, where $r_{\Delta_{\rm c}}$ denotes the radius within which the average mass density of the cluster is $\Delta_{\rm c}$ times the critical density of the Universe at the cluster redshift. For A2255, $\rtwo = 2033$ kpc and $\rfive = 1196$ kpc \citep{ettori19}. \\
\indent
Compared to our previous full-resolution 72~h image published in \citet{botteon22a2255}, the new 224~h image shown in Fig.~\ref{fig:robust-0.5} achieves approximately a factor of 2 better sensitivity, effectively becoming the new deepest radio image ever obtained on a galaxy cluster. This enhanced sensitivity is not solely due to the increased integration time, but also reflects the significant progress in calibration techniques developed over the past years. In particular, whereas the previous image relied only on direction-independent calibration after the extraction step, the present dataset has been corrected for direction-dependent effects across the extracted field, resulting in a more accurate calibration and improved image fidelity. \\
\indent
The radio emission in the field of A2255 is dominated by the extended synchrotron radiation from the ICM and from the cluster active member galaxies. In Fig.~\ref{fig:taper15}, we label the main sources in the field following the nomenclature adopted in previous work \citep{harris80eight, pizzo08, botteon20a2255, botteon22a2255}. In addition to the structures already known, we have labeled a new diffuse source, located $\sim$2.7 Mpc to the east of the cluster center (approximately at the location of $r_{100}$, as $r_{100} \approx 1.36\times\rtwo \approx 2765$ kpc), as a candidate radio relic (``E relic'' label). This emission is also visible at lower significance in the previous \lofar\ images, but was overlooked. Within this diffuse emission, two compact and brighter components are visible: the northern one is likely associated with a spiral galaxy, while the southern source has the morphology of a classical double-lobed radio galaxy, whose optical counterpart is obscured by a bright foreground star (\cf\ Fig.~\ref{fig:E_relic}). We argue that the surrounding extended emission is a radio relic candidate for the following reasons: (i) the two embedded compact sources do not show any clear morphological connection to the larger-scale diffuse emission, suggesting that they are unrelated, (ii) the position in the cluster outskirts, (iii) the elongation of the emission, (iv) its largest projected linear size of $\sim$700 kpc, and (v) the hint of spectral steepening towards the cluster in our previously published spectral index maps (\citealt{botteon22dr2} and Fig.~\ref{fig:E_relic}). \\
\indent
The central region of A2255 hosts the emission detected at the highest signal-to-noise ratio, including some of the brightest and most complex AGN in the system. Leveraging the Ultra-Deep Field dataset, we produced images at progressively higher angular resolution by assigning increasing weight to the long baselines during the deconvolution process. These images are presented in Fig.~\ref{fig:robust_gallery}. In the rightmost panel, we label the main sources. Our highest angular resolution image, with $3.3\arcsec \times 2.2\arcsec$, best highlights the intricate network of filaments in the cluster center originally reported in \citet{botteon20a2255}. Achieving even higher resolutions, \eg\ to resolve the individual filaments within the tailed radio galaxies, requires the use of the \lofar\ international baselines \citep[see][]{derubeis25a2255, derubeis26}. Our new Ultra-Deep Field data also reveal new faint and thin filaments emerging from the terminal part of the Beaver and to the south of the S spur and the Sidekick, which are most clearly visible in the two panels on the left, and the star-forming galaxies with ram-pressure stripped tails identified in \citet{ignesti23a2255}. \\
\indent
In Fig.~\ref{fig:sub_gallery}, we present images showing an increasingly larger FoV centered on the cluster after subtraction of discrete sources. These images were derived from the robust $-$0.5 full-resolution image and from images produced with additional Gaussian tapers of 10\arcsec\ and 25\arcsec\ that were subsequently smoothed to circular beams of 8\arcsec, 16\arcsec, and 35\arcsec. The majority of the discrete sources in the field were successfully removed from the \uv\ data. Exceptions include the brightest and most extended AGN (Bean, Beaver, Embryo, etc.), which we did not attempt to subtract in the visibility domain, and a number of resolved sources (\eg\ compact double-lobed radio galaxies) that were only partially subtracted, leaving clear residuals in the images. In the lowest-resolution and largest-FoV image, a large-scale noise pattern due to Galactic emission is present. %

\section{Synchrotron intensity gradient}\label{sec:sig}

\subsection{Framework}

The SIG technique belongs to the family of the so-called ``gradient methods'' which use gradients of various observables (such as velocity and intensity from spectroscopic data, as well as synchrotron intensity) to infer the projected orientation of magnetic fields in the interstellar and intracluster media \citep[see \eg][]{lazarian24}. This technique relies on the assumption that energy-dominant \alfven\ and slow MHD turbulence modes cascade energy anisotropically, leading to turbulence eddies that are preferentially elongated along the local magnetic field direction. Such elongation suggests that magnetic field fluctuations are stronger in the direction perpendicular to the magnetic field. Consequently, synchrotron intensity, which inherits these fluctuations, exhibits gradients that are dominated by the perpendicular components and are therefore oriented perpendicular to the magnetic field \citep{lazarian16}. SIG has been developed and validated through turbulent box and cluster numerical simulations \citep[\eg][]{lazarian17, zhang19synergies, carmo20, hu25statistics}, and has recently been observationally applied to galaxy clusters by \citet{hu24clusters}. In the context of galaxy clusters, SIG results have been found to be in very good agreement with those obtained from polarization studies of radio relics. In the central regions, typically characterized by the presence of radio halos, however, the results rely primarily on comparisons with numerical simulations, as independent observational constraints are lacking due to the absence of polarized emission in these regions. In this work, we adopt the same framework and summarize the main concepts below. \\
\indent
In the ICM, turbulence is super-\alfvenic\ (\ie\ the \alfven\ Mach number is $\mach_{\rm A} > 1$). From the injection scale $L_{\rm inj}$, the turbulent cascade follows an isotropic Kolmogorov-like scaling of velocity fluctuations (\ie\ $v_l \propto l^{1/3}$, where $v_l$ is the velocity at scale $l$) down to the so-called \alfven\ scale, defined as $l_{\rm A} = L_{\rm inj}\mach_{\rm A}^{-3}$ \citep{lazarian06}. Below this scale, magnetic tension becomes increasingly important, inducing anisotropy in the turbulent eddies. For typical ICM conditions in cluster cores, the \alfven\ scale is estimated to be $l_{\rm A} \approx 1$--60~kpc \citep[\eg][]{brunetti07turbulence}. While this value may differ in cluster outskirts, the SIG method can be applied without precise knowledge of $l_{\rm A}$, since both above and below the \alfven\ scale the intensity gradients are perpendicular to the magnetic field direction. Below the \alfven\ scale, synchrotron intensity gradients are oriented perpendicular to the magnetic field as a result of the anisotropic cascade of MHD turbulence, which produces eddies elongated along the field lines. Above this scale, the magnetic field becomes dynamically subdominant and is passively advected by turbulent motions. This advection also leads to gradients preferentially oriented perpendicular to the magnetic field. Since synchrotron emissivity depends on both the magnetic field and the cosmic ray density, the latter could, in principle, also contribute to the observed gradients. Nevertheless, the emissivity is expected to correlate with the magnetic field in most situations, especially in radio halos, where cosmic rays are accelerated and transported by turbulence and magnetic fields. Their spatial distribution is therefore expected to correlate with the local physical conditions, although this effect has not yet been systematically quantified. For a more detailed theoretical discussion of the applicability of SIG in turbulent ICM conditions, we refer the reader to Section 2 of \citet{hu25statistics}. \\
\indent
As polarization, also SIG is subject to the effects of line-of-sight averaging. In the expected scenario for cluster radio halos, where the line-of-sight integration scale (of the order of Mpc) exceeds the magnetic-field coherence scale (typically tens of kpc), turbulence-induced fluctuations are not expected to exhibit a preferential direction and the alignment between SIG and magnetic fields remains statistically stable, even when the line-of-sight integration length increases \citep{hu24clusters}. These turbulence-driven fluctuations play a major role in reducing the synchrotron polarization fraction through Faraday depolarization. However, SIG is immune to depolarization effects and is specifically designed for turbulence-dominated regions, since turbulent fluctuations become anisotropic in the presence of magnetic fields, and SIG exploits this anisotropy to trace the magnetic field orientation. \\
\begin{figure*}
  \centering
  \includegraphics[width=\hsize,trim={0cm 0cm 0cm 0cm},clip,valign=c]{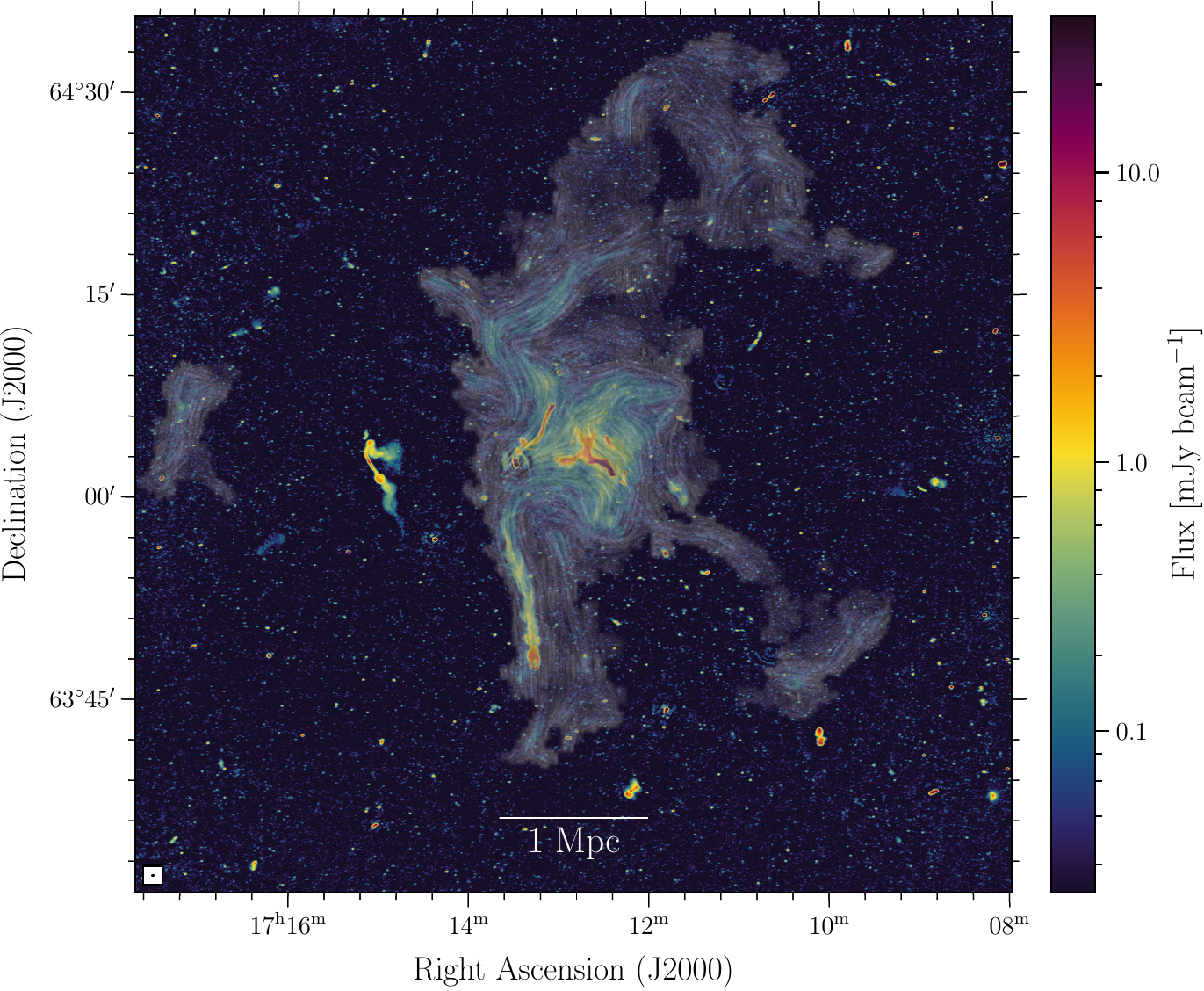}
  \caption{Topology of the magnetic field in A2255. The magnetic field lines inferred by applying the SIG method on the 35\arcsec-resolution image with discrete sources subtracted are visualized on the full-resolution image using the line integral convolution.}
  \label{fig:sig_35_lic}
\end{figure*}
\begin{figure*}
  \centering
  \includegraphics[width=\hsize,trim={0cm 0cm 0cm 0cm},clip,valign=c]{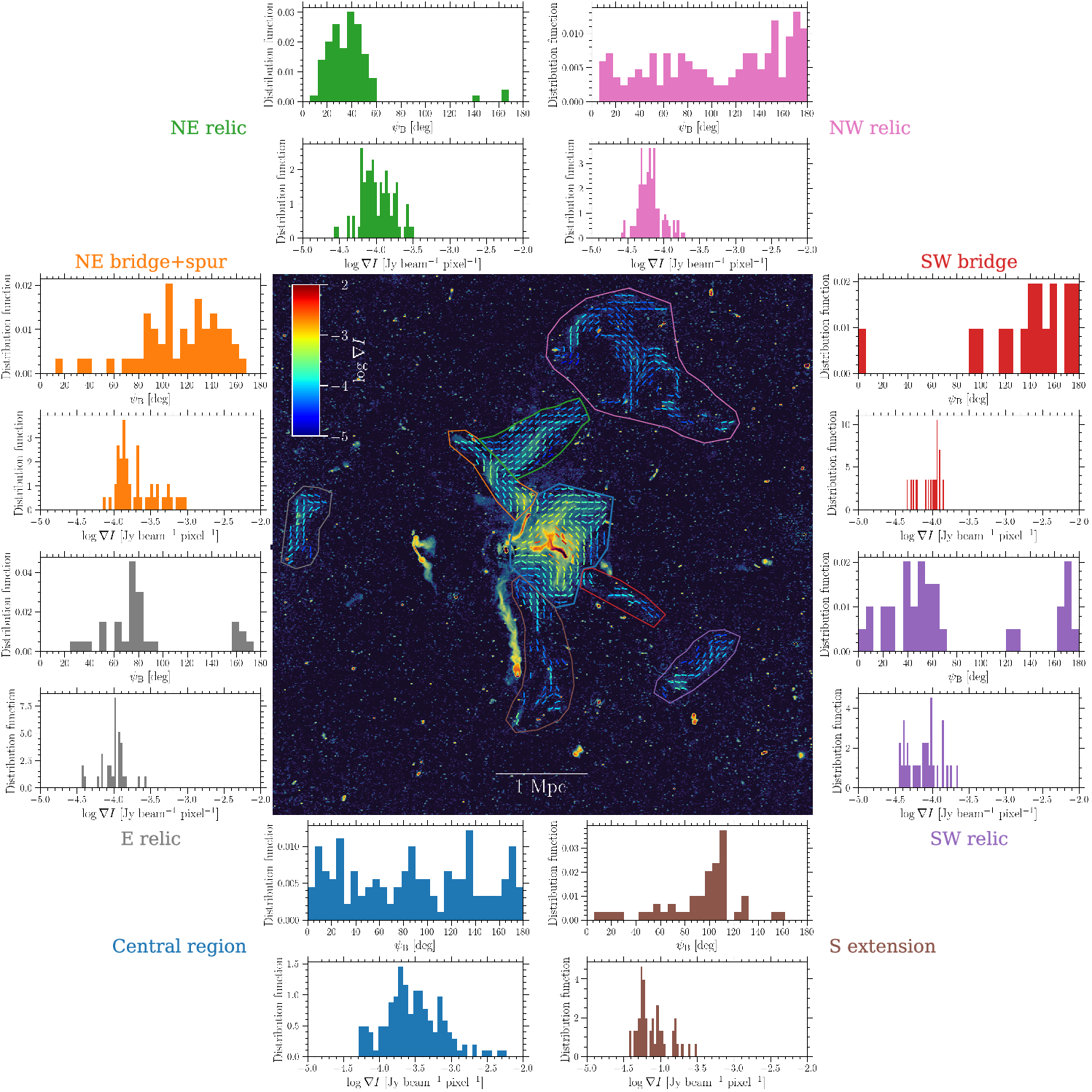}
  \caption{Magnetic fields inferred in specific regions of A2255. Each (magnetic field) segment represents the SIG, color-coded by its amplitude (see colorbar in the figure), averaged for 8 pixel $\times$ 8 pixel for visualization purposes. Histograms show the distribution of gradient angles and amplitudes within the regions defined by the eight polygons.}
  \label{fig:sig_subregions}
\end{figure*}
\begin{figure*}
  \centering
  \begin{minipage}[t]{0.73\linewidth}
  \includegraphics[width=\hsize,trim={0cm 0cm 0cm 0cm},clip,valign=c]{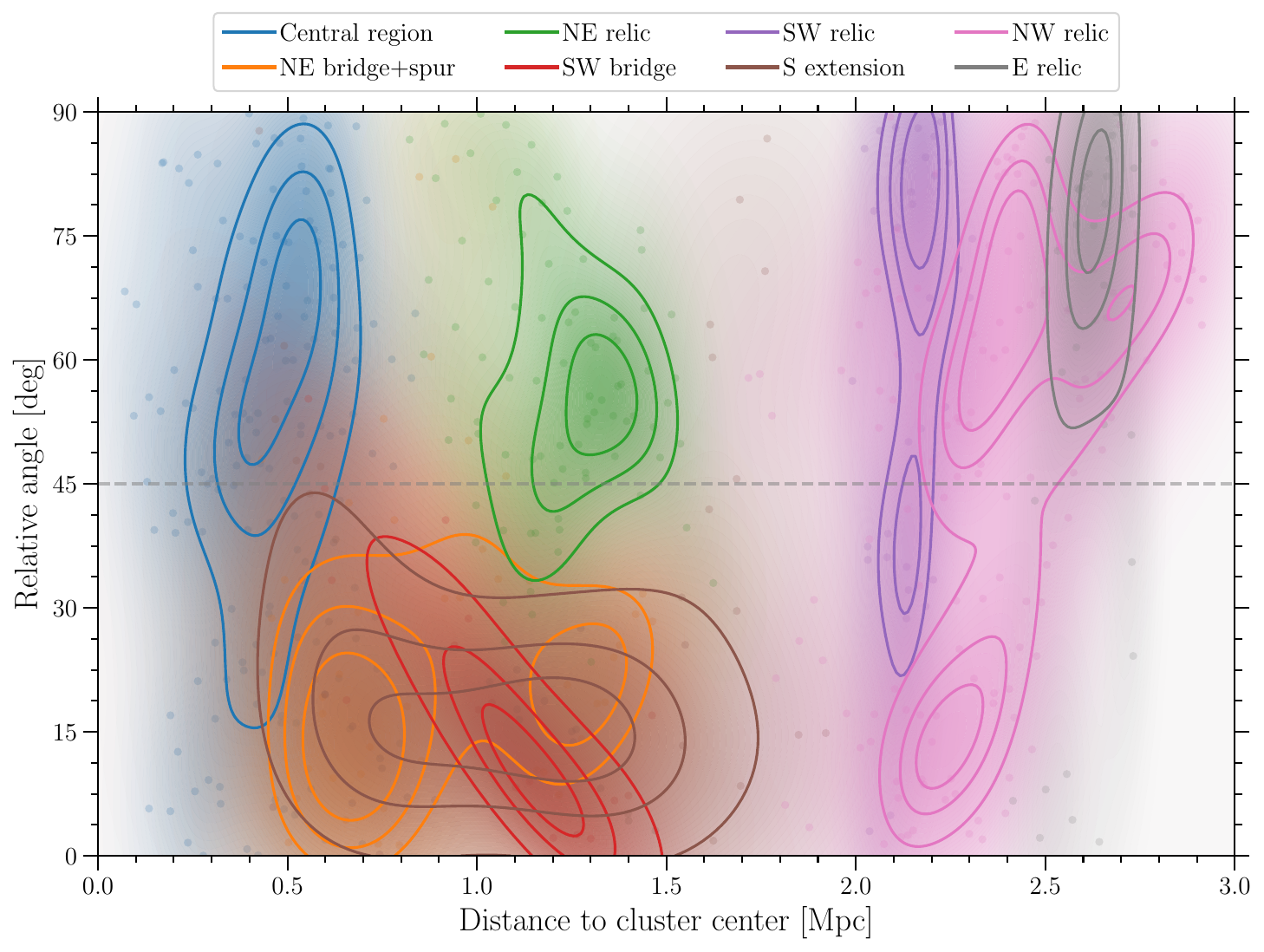}
  \end{minipage}%
  \hfill
  \begin{minipage}[t]{0.25\linewidth}
  \caption{Relative magnetic field orientation ($\Phi$) as a function of projected distance to the cluster center. Different colors denote the eight regions of interest discussed in the paper (see legend). Contours enclose the densest 10\%, 25\%, and 50\% of the distribution computed from the 2D kernel density estimate. A dashed horizontal line is drawn at $\Phi=45\deg$ to separate the regions of the plot where the magnetic field is quasi-radial ($\Phi<45\deg$) and quasi-tangential ($\Phi>45\deg$).}
  \label{fig:DvsPHI}
 \end{minipage}
\end{figure*}
\indent
The SIG method is applied directly to the synchrotron total intensity maps, $I(x,y)$, where the local gradient orientation, $\psi(x,y)$, is computed from finite differences in the two spatial directions:

\begin{equation}\label{eq:gradients}
 \begin{aligned}
 \nabla_x I(x,y)&=I(x+\delta x,y)-I(x,y)/\delta x,\\
 \nabla_y I(x,y)&=I(x,y+\delta y)-I(x,y)/\delta y,\\
 \psi(x,y)&=\tan^{-1}\left(\frac{\nabla_y I(x,y)}{\nabla_x I(x,y)}\right),
 \end{aligned}
\end{equation}

\noindent
where $\nabla_x I(x,y)$ and $\nabla_y I(x,y)$ are the $x$ and $y$ components of the gradient, respectively, and $\delta x$ and $\delta y$ are the scales over which the gradient is computed: in our case, $\delta x = \delta y = 1$ pixel. We note, however, that since pixels in radio images are correlated on the scale of the synthesized beam, the physical scale of the gradients is effectively set by the beam size. To extract statistically meaningful magnetic field orientations and reduce the impact of noise, only pixels with intensities above the $3\sigma_{\rm rms}$ level are considered and a sub-block averaging approach is employed on raw gradient angles \citep{yuen17}. This consists in fitting the histogram of gradient orientations within each sub-block, in our case defined by 20 pixels $\times$ 20 pixels \citep[see][]{hu24clusters}, with a Gaussian function and taking the peak value as the most probable gradient direction for that region. The resulting smoothed gradient map is denoted with $\psi_{\rm s}(x,y)$. \\
\indent
While each sub-block averaging is independent, magnetic field lines are continuous. Therefore, an additional smoothing step based on pseudo-Stokes parameters ($Q_{\rm g}$ and $U_{\rm g}$) derived from the averaged gradients is introduced to improve coherence:

\begin{equation}\label{eq:pseudostokes}
 \begin{aligned}
 Q_{\rm g}(x,y) &= I(x,y)\cos\left[2\psi_{\rm s}(x,y)\right],\\
 U_{\rm g}(x,y) &= I(x,y)\sin\left[2\psi_{\rm s}(x,y)\right],\\
 \psi_{\rm g}(x,y) &= \frac{1}{2}\tan^{-1}\left(\frac{U_{\rm g}}{Q_{\rm g}}\right),
 \end{aligned}
\end{equation}

\noindent
where $\psi_{\rm g}(x,y)$ is the pseudo-polarization angle. The pseudo-Stokes parameters are smoothed with a Gaussian kernel whose full width at half maximum equals the adopted sub-block size. This procedure is analogous to standard treatments of polarized emission and ensures that regions with higher synchrotron intensity contribute more strongly to the inferred field orientation. The final projected magnetic field direction inferred with the SIG method is then obtained as

\begin{equation}\label{eq:psi_B}
 \psi_{\rm B}(x,y) = \psi_{\rm g}(x,y) + \pi/2,
\end{equation}

\noindent
reflecting the perpendicular alignment between synchrotron gradients and magnetic field lines expected in the theoretical framework outlined above. Because SIG cannot probe scales smaller than the beam size, the method is sensitive only to the large-scale orientation of the magnetic field.

\subsection{Application to A2255}\label{sec:sig_a2255}

As our goal is to map the topology of the large-scale diffuse emission in A2255, we applied the SIG technique to the 35\arcsec-resolution image with discrete sources subtracted. The use of an image with a circular beam ensures that the measured gradients are not biased by beam ellipticity. The 35\arcsec\ resolution ($\simeq$53 kpc) represents a good compromise between maximizing the sensitivity to the extended cluster emission and preserving sufficient spatial information to trace magnetic field structures without excessive smoothing. \\
\indent
The magnetic field topology inferred from the SIG analysis is shown in Fig.~\ref{fig:sig_35_lic}. The field orientation is visualized through line integral convolution, overlaid on the full-resolution image. The reconstructed magnetic field lines display coherent and preferential orientations in distinct regions of the cluster. For visualization purposes, the bright tailed AGN embedded in the system are not removed in Fig.~\ref{fig:sig_35_lic}, although they dominate the local gradient orientation and do not trace the magnetic field configuration of the ICM, which is analyzed quantitatively in the following. \\
\indent
In Fig.~\ref{fig:sig_subregions}, we present the analysis of the SIG results, showing the distribution of magnetic field angles, $\psi_{\rm B}$, and gradient amplitudes, $\nabla I$, in eight selected regions of A2255. For this analysis, emission unrelated to the ICM and not subtracted in the \uv\ plane (\ie\ Beaver, Goldfish, Double, Trail, Original TRG, T-bone, Ghost, Sidekick) was blanked, as it does not probe the cluster magnetic field under investigation. The selected regions include the cluster center, the multiple relics, and the emission extensions from the halo, such as the radio bridges. We note that $\psi_{\rm B}$ is defined in Cartesian coordinates, measured from west to north, and we do not distinguish between 0\deg\ and 180\deg. Therefore, an angle of 90\deg\ corresponds to a magnetic field aligned along the north-south direction, while 0\deg\ (or 180\deg) indicates an east-west alignment. As anticipated and now clearly shown by the histograms in Fig.~\ref{fig:sig_subregions}, the inferred magnetic field orientation exhibits preferred alignments in specific regions. In the central region, the angle distribution instead appears significantly broader while the gradient amplitudes take the highest values. This behavior is not unexpected, given the intrinsic morphology of the emission in this region, which is characterized by the presence of numerous surface brightness enhancements and decrements detected at high S/N that contribute to a more complex and less ordered gradient distribution. Our global SIG results do not significantly depend on the inner \uv\ cuts adopted to subtract discrete sources or recover the extended emission, as discussed in Appendix~\ref{app:sig}. \\
\indent
The presence of large-scale diffuse radio emission in A2255 allows us to probe the projected magnetic field orientation from the cluster center out to its outskirts. Since the synchrotron emission from the ICM is likely to be associated with the dissipation of kinetic energy from shocks and turbulence injected by cluster merging activity into nonthermal components, the magnetic field orientation is expected to primarily reflect the local physical processes shaping each emission region. In particular, compression at shocks is expected to produce preferentially tangential field alignments, as in radio relics, while stretching associated with matter infall may lead to more radial configurations, as in filamentary or bridge-like structures. To explore these effects, we examined the relative angle between the magnetic field inferred from the SIG and the radial direction from the cluster center. The distribution as a function of distance is used here only as a convenient way to separate and visualize different emitting regions, rather than to imply an intrinsic dependence of the field orientation on cluster-centric radius. In detail, we proceed as follows: we define the radial angle from the cluster center to a pixel $(x, y)$ as

\begin{equation}\label{eq:radial_angle}
 \phi(x, y) = \tan^{-1} \left( \frac{y-y_{\rm c}}{x-x_{\rm c}} \right),
\end{equation}

\noindent
where $(x_{\rm c}, y_{\rm c})$ denotes the cluster center in pixel coordinates (corresponding to the sky coordinates RA: 17$^{\rm h}$12$^{\rm m}$50.04$^{\rm s}$; DEC: +64\deg03\arcmin10.6\arcsec, \ie\ the center of the image shown Fig.~\ref{fig:sig_subregions}). The absolute difference between the magnetic field orientation and the radial direction is

\begin{equation}\label{eq:ra_difference}
 \Delta(x, y) = |\psi_{B}(x, y) - \phi(x, y)|,
\end{equation}

\noindent
where $\psi_B(x, y)$ is the magnetic field angle at pixel $(x, y)$ as defined in Eq.~\ref{eq:psi_B}. Since orientations separated by 180\deg\ are equivalent, we define the relative radial angle as the smallest angle between the two directions

\begin{equation}\label{eq:phi_abs}
 \Phi(x, y) = \min [(\Delta(x, y), \pi-\Delta(x, y)],
\end{equation}

\noindent
implying that $\Phi = 0\deg$ and $\Phi = 90\deg$ indicate purely radial and tangential magnetic field orientations, respectively. \\
\begin{figure*}[t]
  \centering
  \includegraphics[width=.33\hsize,trim={0cm 0cm 0cm 0cm},clip,valign=c]{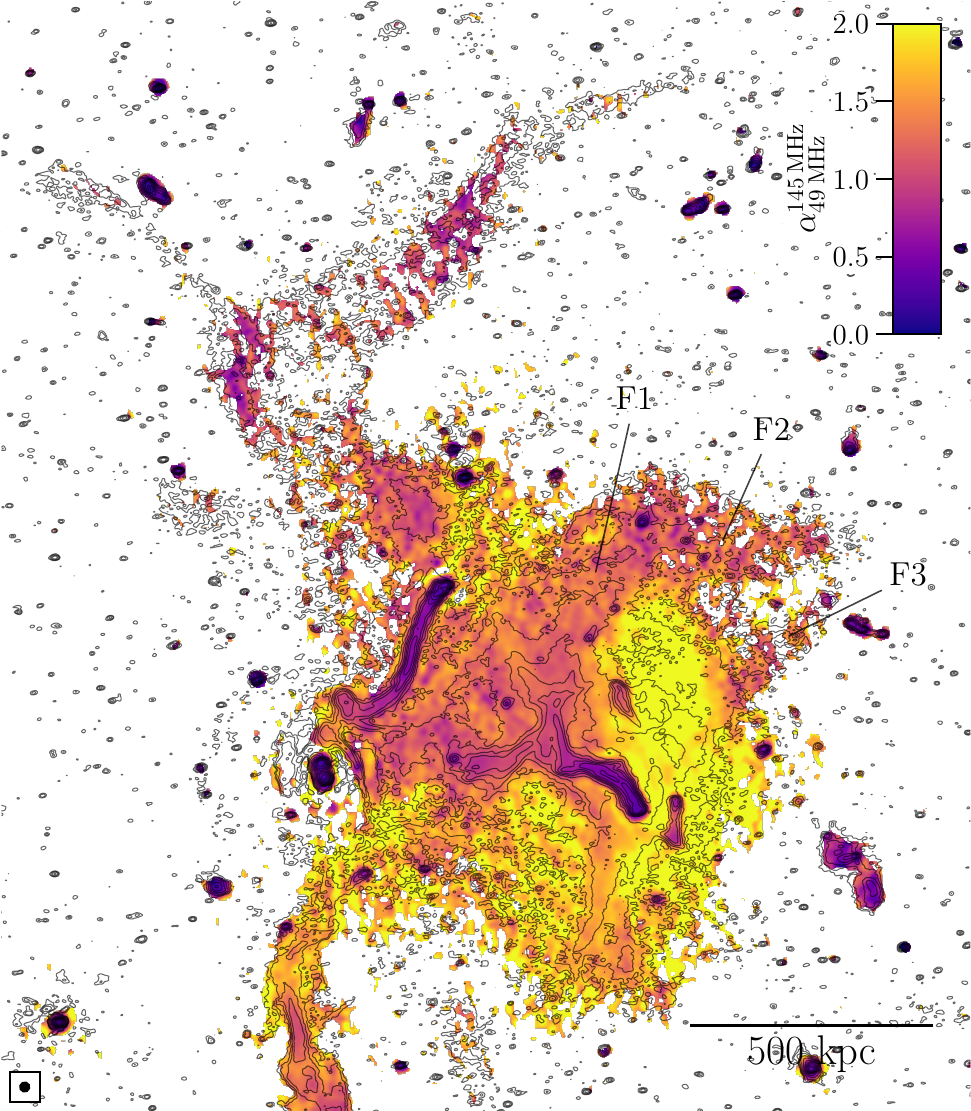}
  \includegraphics[width=.33\hsize,trim={0cm 0cm 0cm 0cm},clip,valign=c]{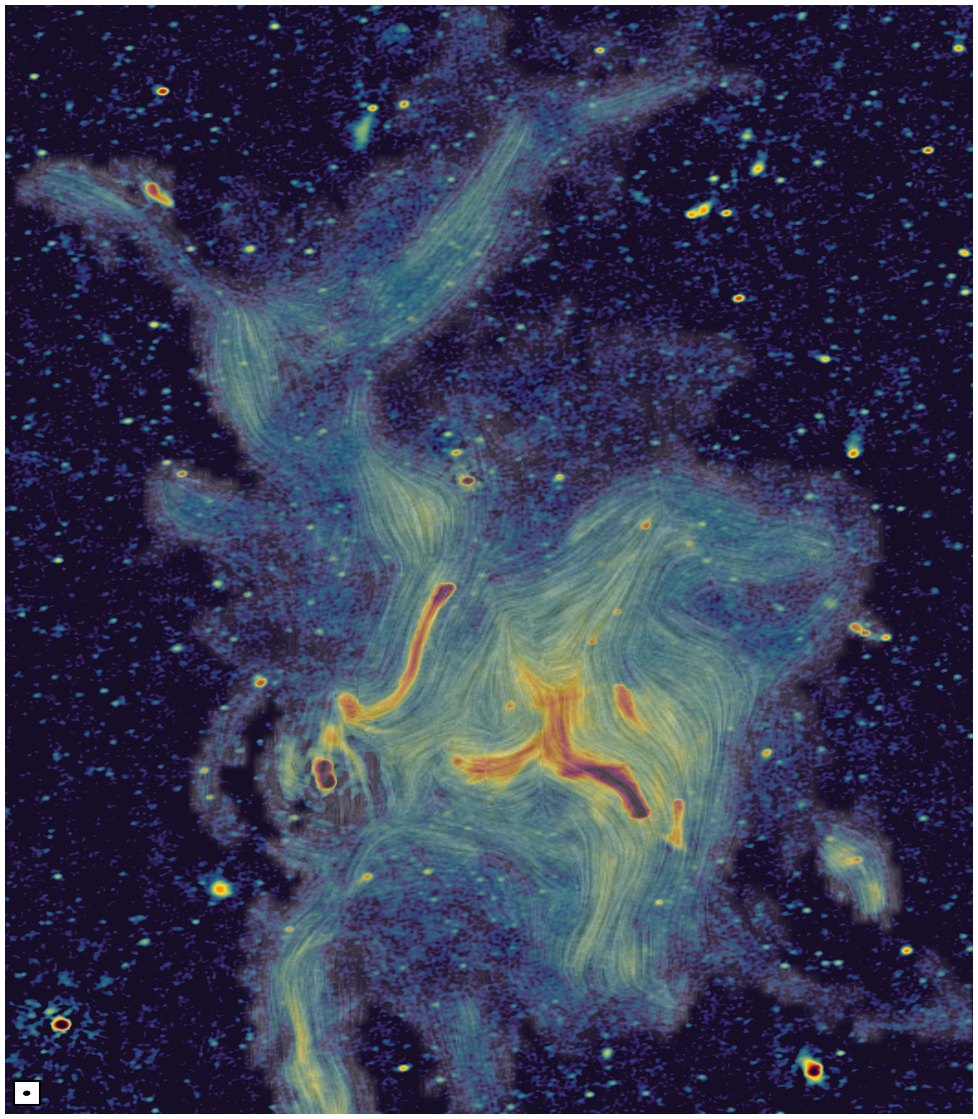}
  \includegraphics[width=.33\hsize,trim={0cm 0cm 0cm 0cm},clip,valign=c]{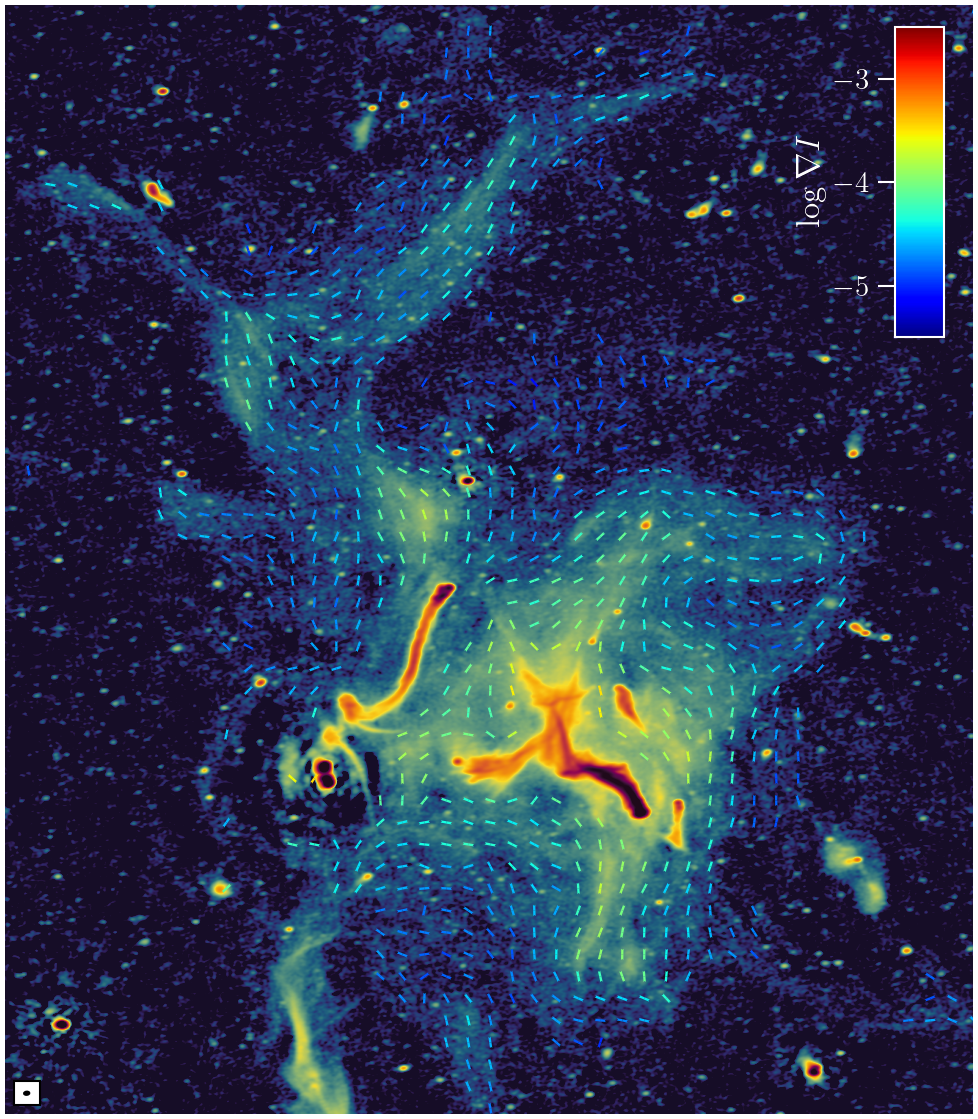}
  \caption{Mapping the central region of A2255. Spectral index ($S_\nu \propto \nu^{-\alpha}$ convention) map between 49 and 145 MHz at 12.5\arcsec\ resolution from \citet{botteon22a2255}, with contours from the Ultra-Deep Field full-resolution image shown in Fig.~\ref{fig:robust-0.5} starting at $3\sigma_{\rm rms}$ (\textit{left panel}). Magnetic field lines obtained by applying the SIG method on the 16\arcsec-resolution image with discrete sources subtracted represented as line integral convolution (\textit{central panel}) and as segments averaged over 12 pixel $\times$ 12 pixel and color-coded by their amplitude (\textit{right panel}). In both panels, the inferred magnetic field lines are visualized on the full-resolution image.}
  \label{fig:halo_relic}
\end{figure*}
\indent
By computing $\Phi(x, y)$ and the on-sky separation between $(x, y)$ and $(x_{\rm c}, y_{\rm c})$, we produced the plot reported in Fig.~\ref{fig:DvsPHI} showing the relative magnetic field orientation obtained with the SIG method as a function of projected distance to the cluster center. The different regions of interest of the cluster are indicated with different colors, while the contours outline the areas of highest probability derived from the 2D kernel density distribution. From the figure, the following distinct behaviors of the inferred magnetic field configuration emerge:

\begin{itemize}
 \item The S extension, NW bridge, and NE bridge+spur typically exhibit quasi-radial fields, being located in the region of the plot with $\Phi<45\deg$.
 \item The field orientation in the NE, SW, NW, and (candidate) E relics is predominantly quasi-tangential (\ie\ $\Phi>45\deg$). We note the complex and broader distribution of angles in the NW relic (already observed in Fig.~\ref{fig:sig_subregions}), reflecting its extended morphology characterized by a bright leading edge and a diffuse, halo-like, component. This combination produces a significant spread in orientations, including an additional peak in the quasi-radial region.
 \item For the central region, the bulk of the magnetic field orientations also seems quasi-tangential, with a tail of the 2D kernel density distribution extending into the quasi-radial region.
\end{itemize}

\noindent
The implications of these findings are discussed next.

\section{Discussion}\label{sec:discussion}

\begin{figure}
  \centering
  \includegraphics[width=0.49\hsize,trim={0cm 0cm 0cm 0cm},clip,valign=c]{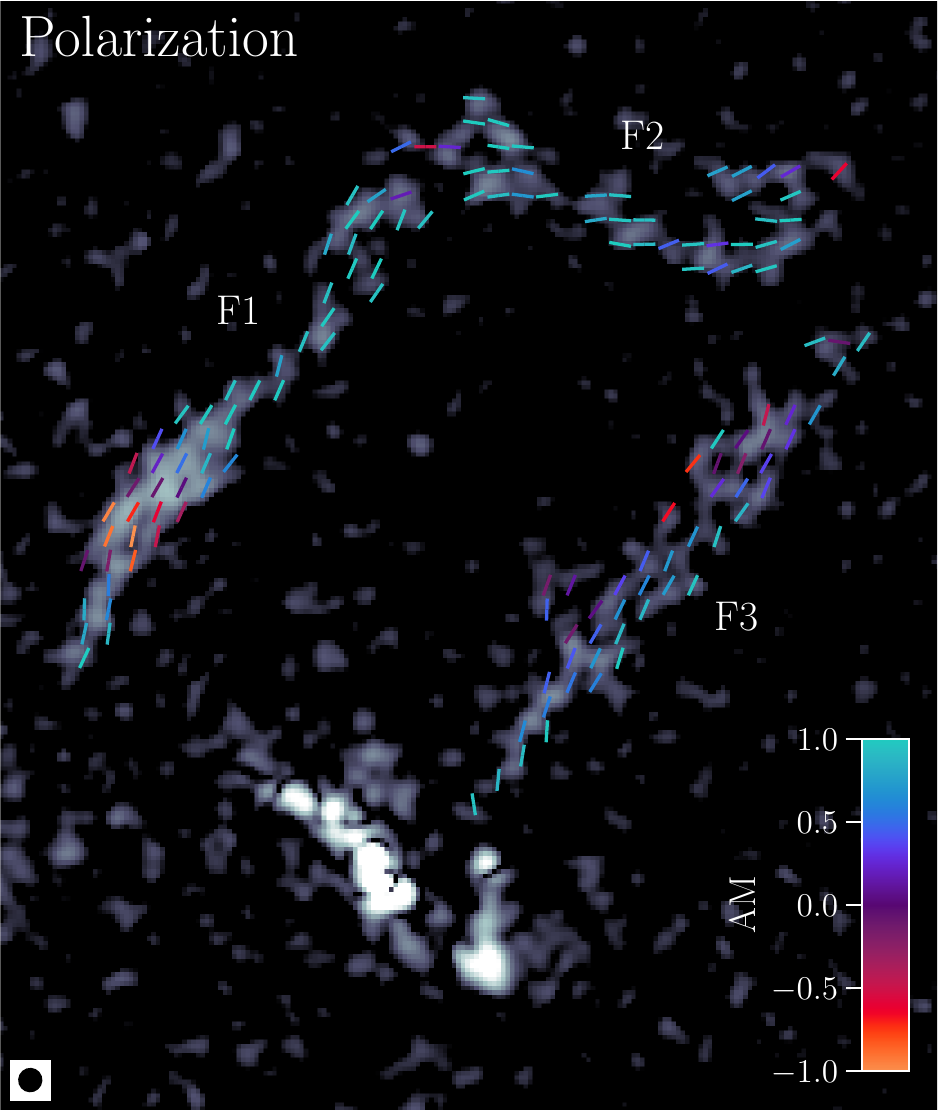}
  \includegraphics[width=0.49\hsize,trim={0cm 0cm 0cm 0cm},clip,valign=c]{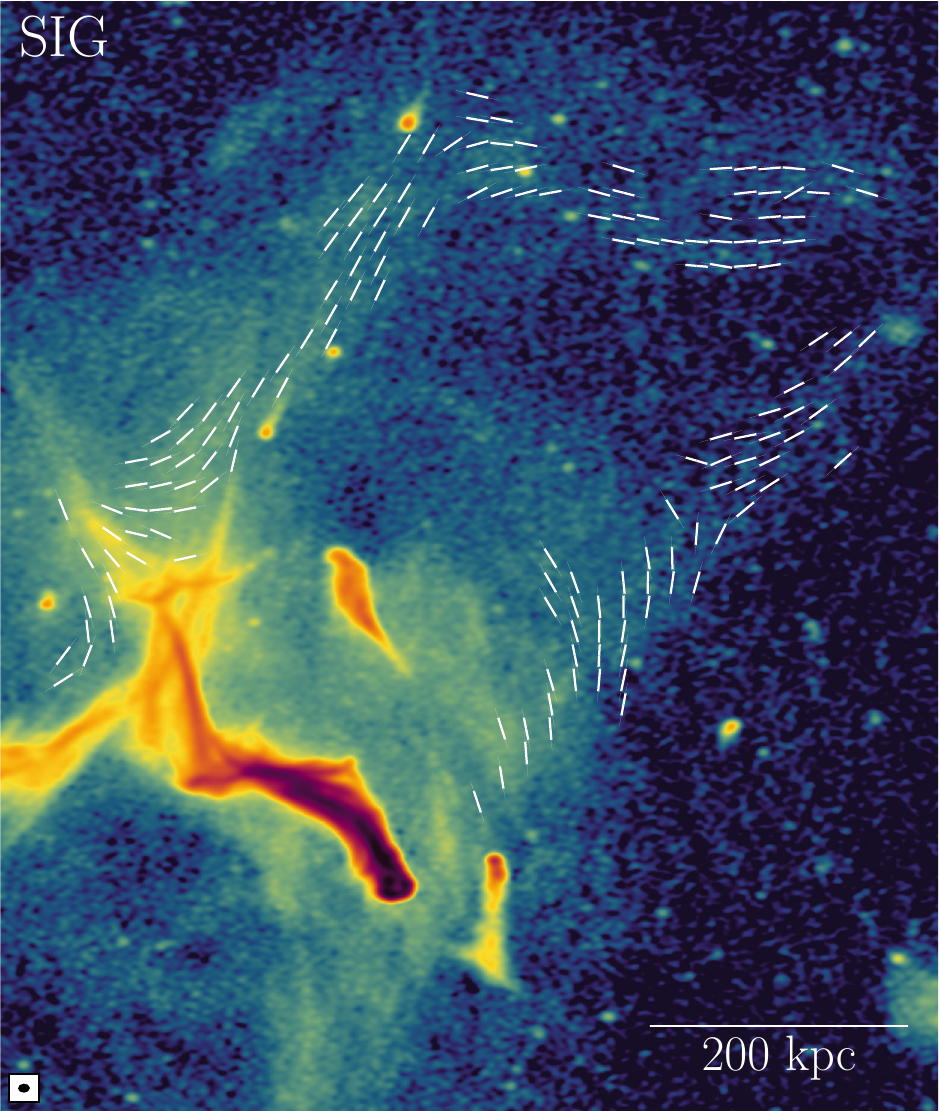}
  \caption{Comparison between the magnetic field orientation inferred from polarization (\textit{left panel}) and from the SIG analysis (\textit{right panel}) for F1, F2, and F3. Magnetic field segments derived from polarization, corrected for Galactic Faraday rotation, are overlaid on the \vla\ 1--2 GHz linearly polarized intensity image at 12\arcsec-resolution (Rajpurohit et al., in prep.) and are color-coded by the AM between SIG and polarization. Magnetic field segments derived from the SIG analysis are overlaid on the image with robust weighting of $-1.0$ shown in Fig.~\ref{fig:robust_gallery}.}
  \label{fig:pol_vs_sig}
\end{figure}

The Ultra-Deep Field observations of A2255 presented in this paper provide the deepest low-frequency radio view of a galaxy cluster obtained to date. One of the primary goal for such a deep observation was to investigate whether the diffuse synchrotron emission in the cluster extends beyond what has been detected in our earlier work \citep{botteon22a2255}. While the Ultra-Deep Field data allowed us to detect the known diffuse structures with higher significance and to recover fainter features within them, the overall spatial extent of the emission is broadly consistent with what previously reported, expanding out to the cluster virial radius but not extending significantly beyond it (with the exception of a portion of the NW relic and the candidate E relic that reach a distance of $\sim$$r_{100}$). This suggests that the emission detected in A2255 may trace the physical boundary of the region where relativistic particles and magnetic fields are sufficiently amplified to produce observable synchrotron radiation. Importantly, by detecting the diffuse emission at higher significance, we were able to probe its structure without the need to employ very low-resolution imaging, which would otherwise smooth out the underlying features and is more prone to artifacts due to calibration and source subtraction residuals. As the sensitivity and resolution of the image reported in Fig.~\ref{fig:robust-0.5} are comparable to the anticipated performance of \ska-Low in its final array configuration, AA4, with 512 stations \citep{braun19arx}, \ska-Low should be able to efficiently probe the complexity of non-thermal phenomena in galaxy clusters across a broad range of spatial scales with only a few hours of integration time. The present \lofar\ Galaxy Cluster Ultra-Deep Field dataset thus provides a glimpse of the observational regime that is expected to become routine in the near future when \ska\ will be fully operational. \\
\indent
Thanks to the SIG analysis performed in this paper, we investigated for the first time the orientation of the magnetic field underlying the large-scale diffuse synchrotron emission of a galaxy cluster. Our results reveal that the inferred magnetic field orientation in A2255 is not random but it exhibits preferential alignments in localized regions of the cluster (Figs.~\ref{fig:sig_subregions} and \ref{fig:DvsPHI}). In particular, the field appears to follow coherent structures in the extensions of the emission from the halo (such as the bridges) and in the relics, preferentially showing quasi-radial and quasi-tangential field orientations, respectively. The central region instead shows a broader distribution of field directions, with the bulk of the SIG segments oriented quasi-tangentially. The origin of this preferential orientation is not immediately clear, but it may be related to the complex dynamics in the central region of the ICM, where the emission exhibits numerous surface brightness features such as decrements, enhancements, edges, and filamentary structures. In this respect, in Fig.~\ref{fig:halo_relic} we provide a higher resolution view of the inferred magnetic fields in this central region, reporting also the 12.5\arcsec-resolution spectral index map of the emission between 49 and 145 MHz obtained by \citet{botteon22a2255}. Here we highlight that the field is strongly aligned along the S spur and the three filaments, labeled F1, F2, and F3 following \citet{pizzo11}, which show flatter spectral index values. These filamentary regions are also polarized (\citealt{govoni05,pizzo11}; Rajpurohit et al., in prep.) and are therefore consistent with being shock-related features projected onto the cluster \citep[see also][]{botteon20a2255}, where the magnetic field is expected to have a preferential orientation. \\
\indent
In Fig.~\ref{fig:pol_vs_sig}, we compare the magnetic field orientation in these three filaments as obtained from our polarization analysis (Rajpurohit et al., in prep.) and as inferred with the SIG method. The SIG results were filtered using a mask that traces the emission of F1, F2, and F3 that is detected in polarization with the \vla\ at 1--2 GHz to allow for a more direct comparison. Polarization vectors are color-coded by the alignment measure (AM, see Eq.~\ref{eq:am}) between the two methods, and indicate an overall good agreement between the field orientation inferred from SIG and polarization, with a median AM value of 0.75. A clear mismatch between the two methods is found for the lower portion of F1: this is due to the fact that the emission from the ``T-bone'' structure, associated with cluster radio galaxies, dominates the orientation of the intensity gradients probed with \lofar\ in this region owing to its high surface brightness, thereby locally affecting the magnetic field orientation inferred with SIG. Interpreting the SIG vectors in this area as tracing the ICM magnetic field is therefore not correct. This region was indeed masked during the SIG analysis presented throughout the rest of the paper, but it was retained in Fig.~\ref{fig:pol_vs_sig} because the mask applied for this comparison was derived from the VLA polarization map, where the ``T-bone'' structure is not detected because of its very steep spectrum. If this region is excluded, the median AM value increases even further. \\
\indent
Future polarization observations will provide an important independent test of the magnetic field orientation inferred from SIG in A2255, particularly for its radio relics. However, several structures, such as the radio halo and the faint features connecting it to the relics, will likely remain undetected in polarization because of Faraday depolarization at low frequencies and their very low surface brightness at higher frequencies. In these regions, SIG may represent the only viable method for probing the large-scale topology of the magnetic field. \\
\begin{figure*}[t]
  \centering
  \includegraphics[width=\hsize,trim={0.4cm 1.5cm 0.75cm 1.5cm},clip,valign=c]{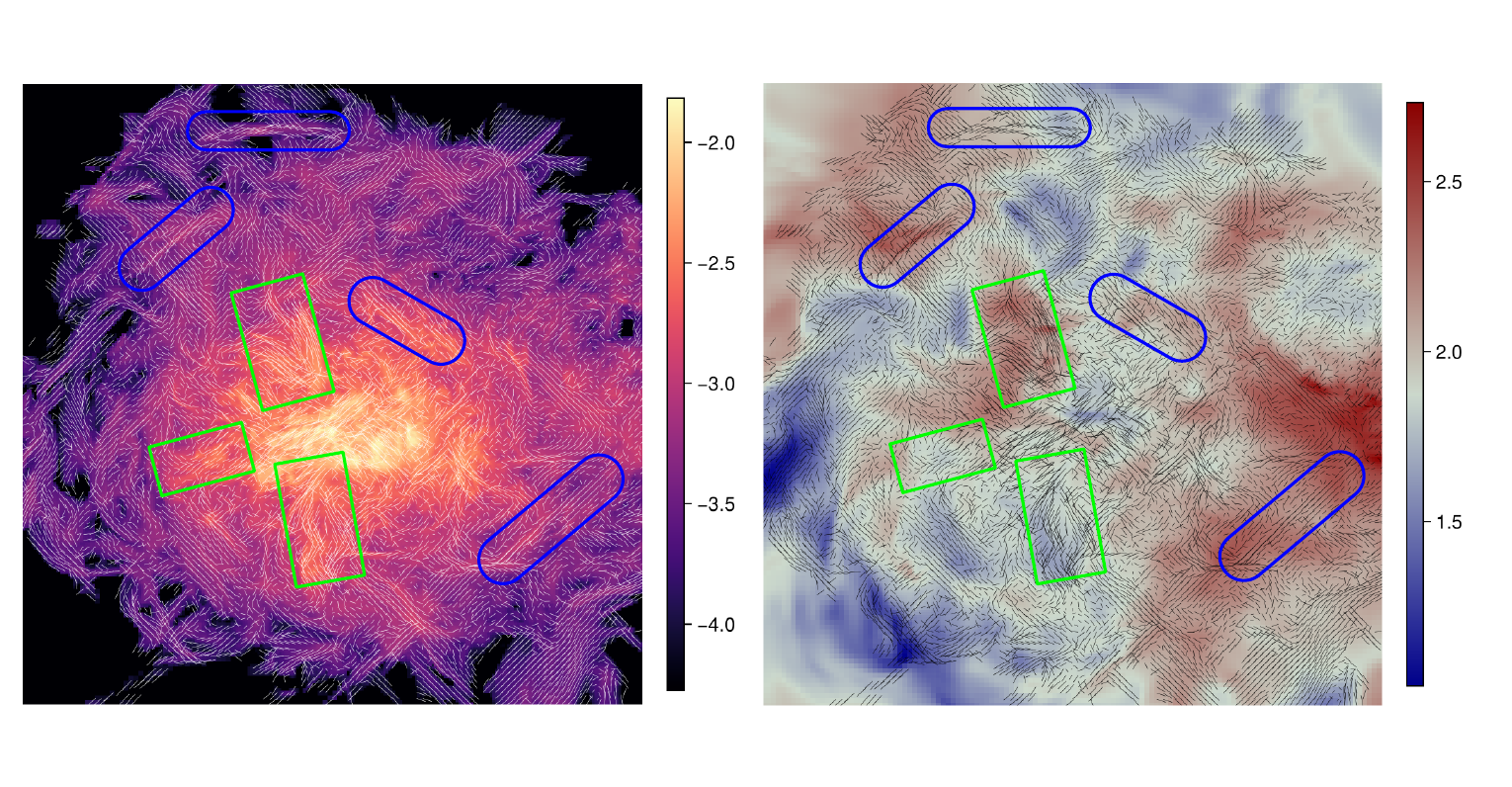}
  \caption{Magnetic field orientation in a cosmological MHD simulation. Segments represent the projected magnetic field along the line-of-sight, weighted by $P_{\rm R}$, and are overlaid on the projected map of $P_{\rm R}$ (\textit{left panel}, logarithmic scale in arbitrary units) and on the projected density-weighted gas velocity field (\textit{right panel}, logarithmic scale of the absolute value in km s$^{-1}$ units). Blue rounded boxes and green rectangles mark regions of shocks and accretion, respectively, where the field orientation has typically tangential and radial orientation. The displayed FoV is 2 Mpc $\times$ 2 Mpc.}
  \label{fig:simulations}
\end{figure*}
\indent
Our findings support a picture in which the topology of the magnetic field in the ICM is closely related to the cluster formation process and the merger dynamics. Cosmological simulations and observational studies have shown that the accretion of matter onto galaxy clusters is not isotropic but it occurs preferentially along large-scale filaments of the cosmic web \citep[\eg][]{bond96, colberg05filaments, springel06}. As a result, radio halos often exhibit protuberances that likely trace the underlying distribution of dark matter and mark regions where the dynamical activity has been enhanced (and thus the radio emission boosted) by the passage of the accreted systems. Since the magnetic field in the ICM is effectively frozen into the plasma, its orientation is expected to follow the large-scale flows associated with the accretion of matter \citep{brzycki19, banfi20, banfi21}. In this scenario, the SE and NW radio bridges in A2255 trace regions where dark matter halos are moving outwards, stretching the magnetic field and producing the quasi-radial orientations observed. At the same time, their motion drives shocks in the cluster outskirts generating the corresponding SE and NW relics, where the magnetic field becomes compressed and ordered, leading to quasi-tangential orientations. Similarly, the S extension may trace a region where enhanced dynamical activity associated with inflowing or outflowing matter increases turbulence in the ICM, enhancing the radio emission and producing a quasi-radial field configuration. We note that the dynamically active nature of the regions around the NW relic and the S extension in A2255 is supported by optical observations, which reveal the presence of a group of galaxies with a significant line-of-sight velocity offset \citep{golovich19atlas} and an elongation of the weak-lensing contours (\citealt{finner25}; K.~Finner, private communication), respectively. Spectroscopic redshifts are essential to accurately separate cluster members from foreground and background galaxies, thereby improving the fidelity of the weak-lensing mass reconstruction. Therefore, a deep spectroscopic survey of galaxies in A2255 extending out to its virial radius would enable the reconstruction of a detailed weak-lensing mass map, providing a direct test of the proposed connection between cluster dynamics, anisotropies of the diffuse emission, and magnetic field orientations. \\
\indent
While the amplification of the ICM magnetic field during cluster mergers has been extensively studied with numerical simulations (\eg\ \citealt{roettiger99magnetic, dolag02, vazza18dynamo, steinwandel24}, and \citealt{donnert18rev} for a review), its topology has received less attention, with most of the studies focusing on the field orientation in clusters experiencing sloshing motions, where shear flows can produce characteristic magnetic layers and ordered field structures \citep[\eg][]{zuhone11sloshing, zuhone21transport, zuhone21bubbles, hu25statistics, lehle26}, and only a few investigating the structure of the magnetic field in cluster mergers and on larger scales \citep{bruggen05, marinacci15, banfi20, banfi21}. The highly sensitive radio observations presented in this work, combined with the SIG method, provide the first attempt to map the projected orientation of the magnetic field across an entire cluster, from core to outskirts. These measurements motivate new MHD cosmological simulations aimed to explore the evolution and topology of the ICM magnetic during the cluster formation process. By following the anisotropic accretion of matter onto clusters, such simulations can help us understand how turbulence and shocks shape the magnetic field on large scales, potentially leading to coherent structures in specific cluster regions, as inferred by the SIG method in A2255. \\
\indent
We perform a first qualitative comparison with state-of-the-art cosmological simulations of ICM magnetic fields by using a recent MHD simulations produced with the \textsc{enzo} code \citep{bryan14}. In particular, we consider the resimulation of the same Coma-like galaxy cluster used in \citet{vazza26}.  The $z=0$  mass of this cluster is close to the one of the real Coma ($M_{200} \approx 1.1 \times 10^{15} M_{\odot}$ at $z=0.02$), it accreted $\sim 50\%$ of its final virial mass during a major merger (i.e. a mass ratio larger than one fifth of the cluster mass at that time) at around $z \sim 0.45$, and the radial profile of its simulated Faraday rotation is consistent with that measured in the Coma cluster \citep[][]{vazza18dynamo}. Also, the turbulence generated in the simulation allows one to reproduce the observed radio halo under the assumption of second-order acceleration of relativistic electrons \citep{bonafede21}, while at the same time producing a mock statistics of X-ray turbulent broadening of iron emission lines compatible with the latest observations by the \xrism\ satellite \citep{vazza26}. This ideal MHD simulation used up to eight levels of adaptive mesh refinement to reach a peak spatial resolution of  $\Delta x = 3.95$ kpc/cell in most of the cluster core, and has otherwise a spatial resolution of $\Delta x= 15.8$ kpc/cell in the entire virial volume, as in  \citet[][]{vazza18dynamo}. However, compared to previous work, this resimulation of the same object used the constrained transport method to ensure that $\nabla \cdot {\bf B}$ is as small as allowed by numerical roundoff errors \citep[\eg][]{collins10}. The initial magnetic seed field (uniformly seeded in the entire simulated volume at the start of the simulation) is $B_0 = 0.3$ nG in all directions. \\
\indent
In Fig.~\ref{fig:simulations} we show the projected distribution of magnetic field vectors along the line-of-sight, weighting each component by a simple proxy for synchrotron emission, \ie\ $P_{\rm R} = n_{\rm e} B^2$, where $n_{\rm e}$ is the electron number density and $B$ is the magnetic field strength in each cell. The vectors are overlaid on the projected map of $P_{\rm R}$ (\textit{left panel}) and on the projected density-weighted gas velocity field (\textit{right panel}), applying a smoothing that mimics the observational procedure \\
\indent
The topology of projected magnetic field vectors is manifestly complex and organized on many scales, which depend on the local gas dynamics. In the figure, we identify a few sample structures, which show some qualitative resemblance to the pattern highlighted by our SIG analysis of A2255. In several cases, we observe magnetic field stretched in a tangential direction with respect to the cluster radius, which we can unambiguously identify (also based on the analysis of thin slices of other quantities, \eg\ gas temperature and the divergence of the velocity field) as weak shock waves triggered by accretion flows. In other cases, magnetic fields are stretched radially to the cluster center, and in these cases we typically find that they are correlated with large gas flows towards the cluster center, driven by the accretion of gas substructures onto the main cluster. Of course, in many other cases (like in the cluster central region) the projected topology of magnetic fields is less trivial to interpret, due to the mixing of different flows and to projection effects along the line-of-sight. \\
\indent
Overall, this test suggest that it is possible to observe coherent patches and spurs of magnetic fields on large-scales (hundreds of kpc) in the ICM. These features do not necessarily correspond to laminar magnetic field configurations with a uniform direction, but rather arise from compression and stretching driven by the local ICM dynamics, which organize the otherwise tangled magnetic field generated by the small-scale dynamo process. This picture is thus not inconsistent with RM studies of clusters, which typically infer magnetic field auto-correlation lengths of a few tens of kpc in the ICM \citep[see \eg][for a recent collection of results]{loi26}.

\section{Conclusions}\label{sec:conclusions}

We have presented the \lofar\ Galaxy Cluster Ultra-Deep Field, in which A2255 ($z=0.080$) has been observed for a total of 336~h in the frequency range 120--168 MHz. Our images employ the 224~h of data with the best quality, which we used to produce deep images at different resolutions covering a field of 4 deg$^2$ centered on the cluster. These observations provide the deepest low-frequency radio view of a galaxy cluster obtained to date. \\
\indent
The improved sensitivity and advanced calibration techniques allowed us to recover the extended emission across the cluster out to its virial radius at higher significance. With the exception of a new candidate radio relic located $\sim$2.7~Mpc to the east of the cluster center, our images do not reveal extended emission on scales significantly larger than previously reported. Instead, the central region of the cluster is better resolved, revealing the complexity of the diffuse emission and cluster radio galaxies, which are characterized by numerous surface brightness structures such as thin synchrotron filaments. \\
\indent
We applied the SIG technique to 35\arcsec-resolution images with discrete sources subtracted to investigate the topology of the ICM magnetic field across the cluster. The results show that the inferred magnetic field orientation is not random but exhibits coherent patterns in different regions: quasi-radial fields are observed in the halo extensions and bridges, while the relics predominantly show quasi-tangential orientations. The central region instead exhibits a broader distribution of field directions, consistent with a turbulent medium. This is in line with a scenario in which the magnetic field is frozen into the ICM and its topology is shaped by the cluster merger dynamics and large-scale accretion flows. This interpretation is qualitatively supported by comparisons with cosmological MHD simulations. \\
\indent
Our findings showcase the potential of combining ultra-deep low-frequency imaging with the SIG technique as a powerful method for investigating the magnetic field in the ICM from cluster centers to the outskirts. This approach is complementary to the traditional Faraday RM analysis typically performed at frequencies $\gtrsim$1 GHz, and is particularly valuable because it is not affected by Faraday depolarization. While the SIG technique probes the large-scale magnetic field structure, RM is sensitive to smaller-scale fluctuations, where the field is more tangled. \\
\indent
The images and results presented here provide a glimpse of what is expected to become routinely accessible with \ska-Low in the near future in terms of the level of details, the recovery of extended emission, and the study of magnetic fields in galaxy clusters. In combination with numerical simulations and complementary multi-wavelength observations, these capabilities will be essential to further probe the processes that shape magnetic fields during cluster formation.

\begin{acknowledgements}
We express our gratitude to Hanno Holties and Arpad Miskolczi for helping optimizing the download of $\sim$200 TB of data from the \lofar\ LTA, and to Francesco Bedosti, Matteo Gandolfi, Gianmarco Maggio, and Giuliano Taffoni for technical support on the \lofar-Pleiadi and HOTCAT HPC systems.
ABot acknowledges the financial contribution from the INAF Mini-Grant 1.05.23.04.01 {\it ``The LOFAR Galaxy Cluster Ultra-Deep Field''}.
FV acknowledges funding under the European Union's  Horizon Europe program through the ERC Synergy Grant COSMOMAG (Project Id. 101224803), and the
CINECA award ``IscrB{\_}CREW'' under the ISCRA initiative, for the availability of high-performance computing resources and support, and the usage of online storage tools kindly provided by the INAF Astronomical Archive (IA2) initiative (\url{http://www.ia2.inaf.it}).
EDR acknowledges support by the Deutsche Forschungsgemeinschaft (DFG).
MB acknowledges the financial contribution from the INAF GO grant 1.05.24.02.10 {\it ``Extended Radio Emission in Galaxy Clusters at deep focus with MeerKAT''}.
ABon and MB acknowledge support from the ERC CoG $\vec{B}$ELOVED, GA N.101169773.
GDG acknowledges support from the ERC Consolidator Grant ULU 10108637.
MJH thanks the UK STFC for support [ST/V000624/1, ST/Y001249/1].
AI acknowledges support from the institutional project RVO:67985815 and the project 25-19512L of the Czech Science Foundation.
This paper is based (in part) on data obtained with the LOFAR telescope (LOFAR-ERIC) under project codes LC12\_027 and LT16\_005. LOFAR \citep{vanhaarlem13} is the LOw Frequency ARray designed and constructed by ASTRON. It has observing, data processing, and data storage facilities in several countries, that are owned by various parties (each with their own funding sources), and that are collectively operated by the LOFAR European Research Infrastructure Consortium (LOFAR-ERIC) under a joint scientific policy. The LOFAR-ERIC resources have benefited from the following recent major funding sources: CNRS-INSU, Observatoire de Paris and Universit\'{e} d'Orl\'{e}ans, France; Istituto Nazionale di Astrofisica (INAF), Italy; BMBF, MIWF-NRW, MPG, Germany; Science Foundation Ireland (SFI), Department of Business, Enterprise and Innovation (DBEI), Ireland; NWO, The Netherlands; The Science and Technology Facilities Council, UK; Ministry of Science and Higher Education, Poland. This research made use of the Dutch national e-infrastructure with support of the SURF Cooperative (e-infra 180169) and the LOFAR e-infra group. The J\"{u}lich LOFAR Long Term Archive and the German LOFAR network are both coordinated and operated by the J\"{u}lich Supercomputing Centre (JSC), and computing resources on the supercomputer JUWELS at JSC were provided by the Gauss Centre for Supercomputing e.V. (grant CHTB00) through the John von Neumann Institute for Computing (NIC). This research made use of the Italian LOFAR-IT computing infrastructure supported and operated by INAF, including the resources within the PLEIADI special ``LOFAR'' project by USC-C of INAF, and by the Physics Department of Turin university (under an agreement with Consorzio Interuniversitario per la Fisica Spaziale) at the C3S Supercomputing Centre, Italy.
This work made use of the following \textsc{python} packages: \texttt{astropy} \citep{astropy22}, \texttt{CMasher} \citep{vandervelden20}, \texttt{matplotlib} \citep{hunter07}, \texttt{numpy} \citep{vanderwalt11}, and \texttt{scipy} \citep{virtanen20}.
\end{acknowledgements}

\bibliographystyle{aa}
\bibliography{library.bib}

\begin{appendix}

\onecolumn

\section{List of observations}\label{app:list}

\begin{table}[h]
\centering
\caption{\lofar\ Galaxy Cluster Ultra-Deep Field observations.}
\label{tab:radio_obs}
\begin{tabular}{rrrrrrrr}
\hline
\hline
 \# & SAS ID & Project code & Observing date & Calibrator used & Data quality \\
\hline
1  & 720376  & LC12\_027 & 2019-06-07 & 3C48    & Medium \\
2  & 725452  & LC12\_027 & 2019-06-22 & 3C48    & Good \\
3  & 726706  & LC12\_027 & 2019-06-28 & 3C48    & Good \\
4  & 727108  & LC12\_027 & 2019-07-03 & 3C48    & Good \\
5  & 728072  & LC12\_027 & 2019-07-08 & 3C295   & Good \\
6  & 733075  & LC12\_027 & 2019-08-09 & 3C48    & Good \\
7  & 746862  & LC12\_027 & 2019-09-28 & 3C295   & Medium \\
8  & 747611  & LC12\_027 & 2019-10-04 & 3C48    & Good \\
9  & 751364  & LC12\_027 & 2019-11-15 & 3C48    & Good \\
10 & 863650  & LT16\_005 & 2022-06-12 & 3C295   & Bad \\ %
11 & 863664  & LT16\_005 & 2022-06-11 & 3C295   & Good \\ %
12 & 863678  & LT16\_005 & 2022-06-23 & 3C295   & Medium \\
13 & 863706  & LT16\_005 & 2022-07-03 & 3C295   & Good \\
14 & 863720  & LT16\_005 & 2022-06-29 & 3C295   & Good \\
15 & 863734  & LT16\_005 & 2022-06-19 & 3C295   & Bad \\
16 & 863748  & LT16\_005 & 2022-06-16 & 3C295   & Good \\
17 & 863762  & LT16\_005 & 2022-07-08 & $\dots$ & Failed \\
18 & 863776  & LT16\_005 & 2022-06-25 & 3C295   & Bad \\
19 & 863790  & LT16\_005 & 2022-07-10 & $\dots$ & Failed\\
20 & 863804  & LT16\_005 & 2022-06-18 & 3C295   & Bad \\
21 & 865600  & LT16\_005 & 2022-07-14 & 3C295   & Medium \\
22 & 865614  & LT16\_005 & 2022-07-17 & 3C295   & Bad \\
23 & 865628  & LT16\_005 & 2022-07-21 & $\dots$ & Failed \\
24 & 865642  & LT16\_005 & 2022-07-22 & 3C295 & Bad \\
25 & 865656  & LT16\_005 & 2022-07-24 & $\dots$ & Failed \\
26 & 865670  & LT16\_005 & 2022-08-18 & 3C295   & Good \\
27 & 2015075 & LT16\_005 & 2023-04-10 & 3C295   & Good \\
28 & 2015093 & LT16\_005 & 2023-04-15 & 3C295   & Good \\
29 & 2015102 & LT16\_005 & 2023-04-16 & 3C295   & Good \\
30 & 2015120 & LT16\_005 & 2023-04-20 & 3C295   & Good \\
31 & 2015129 & LT16\_005 & 2023-04-21 & 3C295   & Good \\
32 & 2015138 & LT16\_005 & 2023-04-22 & 3C295   & Good \\
33 & 2015147 & LT16\_005 & 2023-04-23 & 3C295   & Good \\
34 & 2015156 & LT16\_005 & 2023-04-24 & 3C295   & Medium \\
35 & 2015165 & LT16\_005 & 2023-04-27 & 3C295   & Medium \\
36 & 2015174 & LT16\_005 & 2023-04-28 & 3C295   & Bad \\
37 & 2015183 & LT16\_005 & 2023-04-29 & 3C295   & Medium \\
38 & 2015192 & LT16\_005 & 2023-05-04 & 3C295   & Good \\
39 & 2015201 & LT16\_005 & 2023-05-03 & 3C295   & Medium \\
40 & 2015210 & LT16\_005 & 2023-04-30 & 3C295   & Bad \\
41 & 2017002 & LT16\_005 & 2023-05-01 & 3C295   & Bad \\
42 & 2017156 & LT16\_005 & 2023-05-08 & 3C295   & Bad \\
\hline
\end{tabular}
\end{table}

\section{Impact of short baselines on the modeling of discrete sources and recovering of the extended emission}\label{app:systematicunc}

\begin{figure*}
  \centering
  \includegraphics[width=.94\hsize,trim={0cm 0cm 0cm 0cm},clip,valign=c]{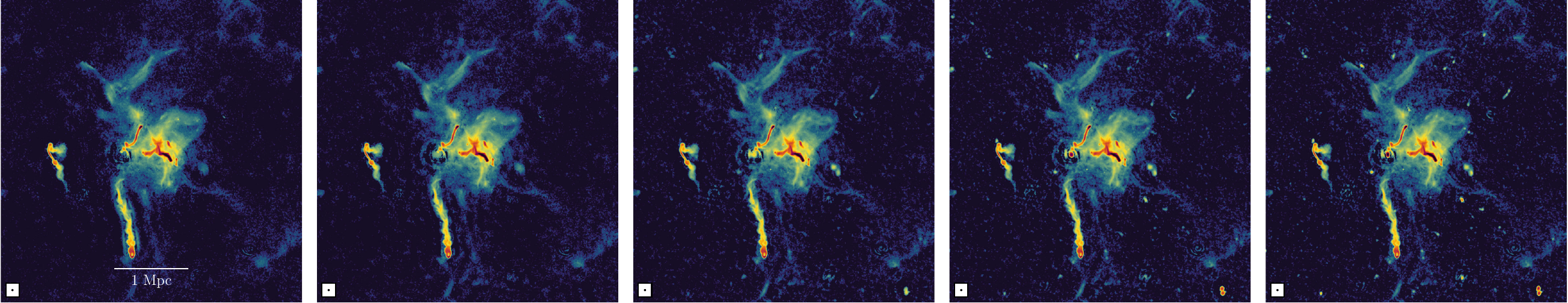}
  \includegraphics[width=.94\hsize,trim={0cm 0cm 0cm 0cm},clip,valign=c]{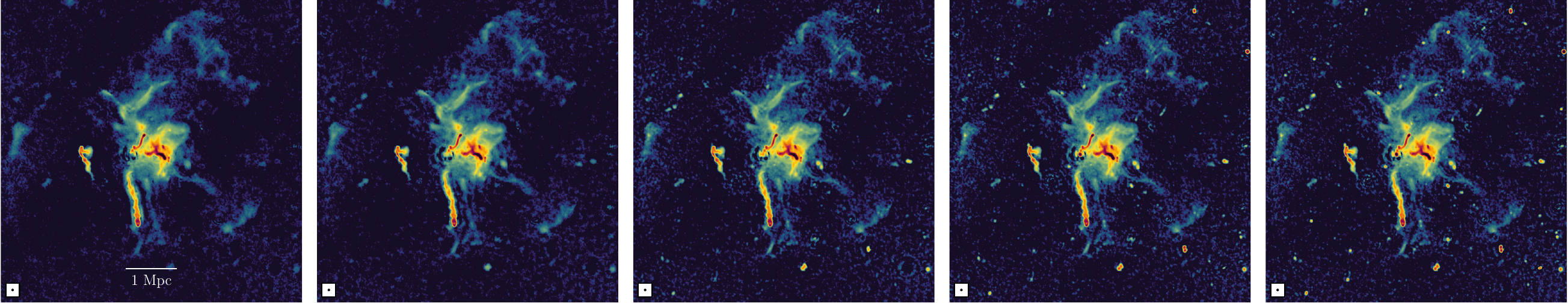}
  \includegraphics[width=.94\hsize,trim={0cm 0cm 0cm 0cm},clip,valign=c]{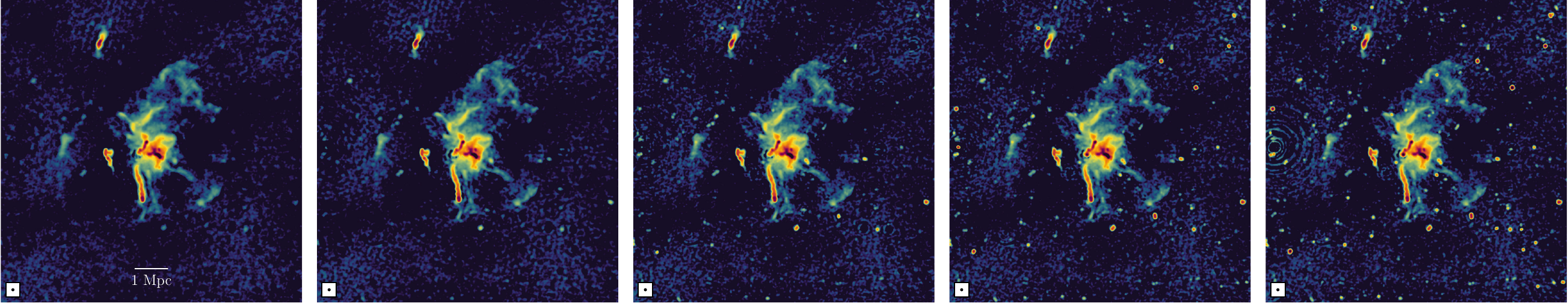}
  \caption{Impact of the choice of the inner \uv\ cut for discrete source subtraction on the final images. From \textit{top} to \textit{bottom}, images were smoothed to circular beams of 8\arcsec, 16\arcsec, and 35\arcsec. From \textit{left} to \textit{right}, each panel corresponds to discrete source subtracted images modeled using inner \uv\ cuts of 1500, 2500, 5000, 8000, and 10000$\lambda$. The three images in the central column correspond to those shown in Fig.~\ref{fig:sub_gallery}. The radio beams are shown in the bottom left corners.}
  \label{fig:subtest_gallery}
\end{figure*}

\begin{figure*}
  \centering
  \includegraphics[width=.94\hsize,trim={0cm 0cm 0cm 0cm},clip,valign=c]{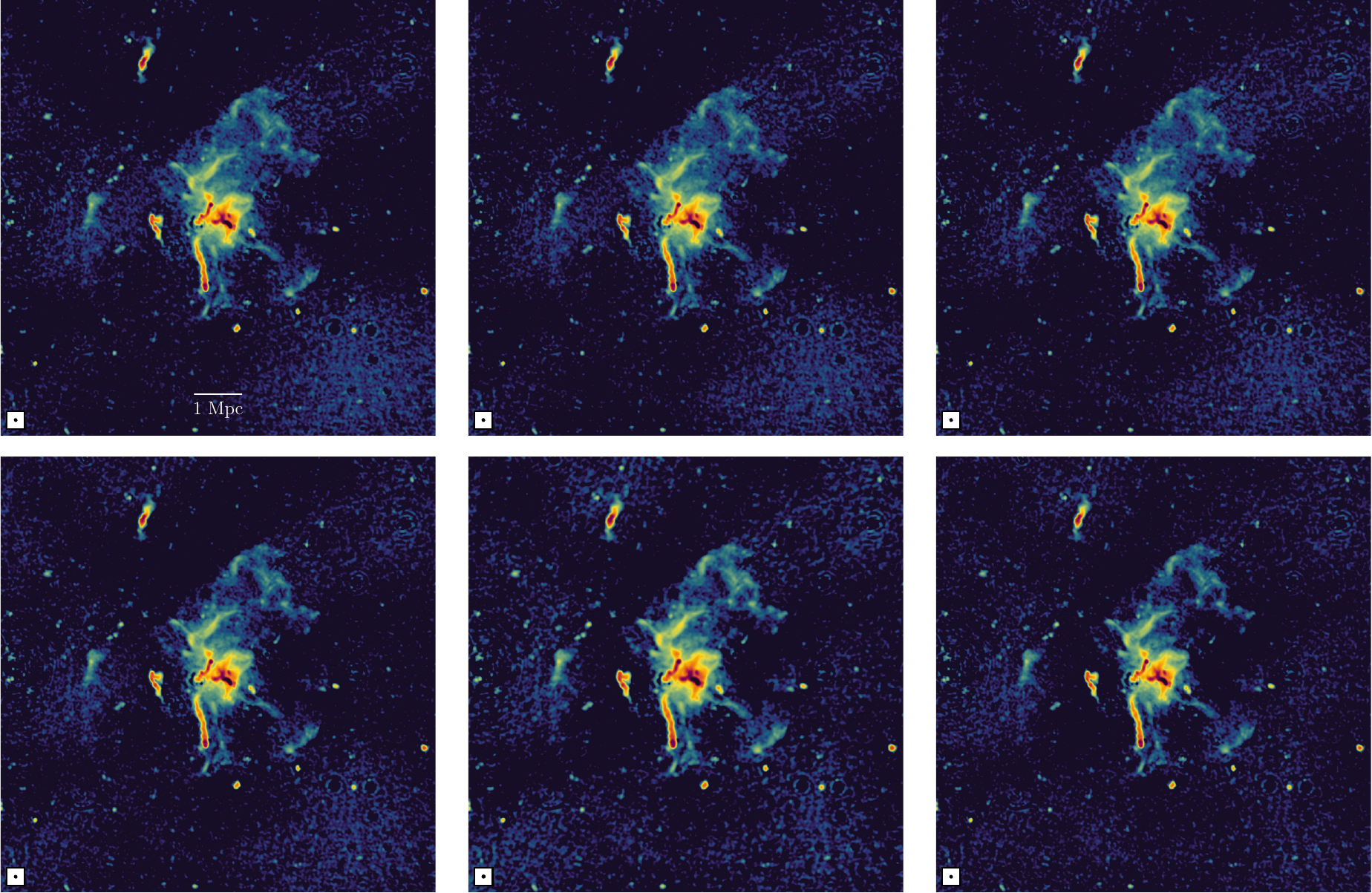}
  \caption{Impact of the choice of the inner \uv\ cut for the recovery of extended emission on the final images. From \textit{top left} to \textit{bottom right}, images with a circular beam of 35\arcsec\ produced using inner \uv\ cuts of 20, 30, 40, 50, 60, and 70$\lambda$. The image in the central column of the bottom row corresponds to the rightmost panel shown in Fig.~\ref{fig:sub_gallery}. The radio beams are shown in the bottom left corners.}
  \label{fig:shortbaselinestest_gallery}
\end{figure*}

\subsection{Images}\label{app:images}

In order to subtract the contribution of discrete sources from the visibility data, a model of discrete sources was obtained by applying an inner \uv\ cut during imaging. The choice of the inner \uv\ cut depends on the maximum angular scale of the emission that we aim to subtract, as emission on larger scales is filtered out by the interferometer. On the one hand, adopting a large inner \uv\ cut allows to filter out the extended emission from the ICM which we aim to preserve during the modeling process. On the other hand, excessively large cuts may prevent us from capturing the extended emission associated with discrete sources that we aim to subtract. The choice of the inner \uv\ cut is therefore a trade-off between these competing effects. In Fig.~\ref{fig:subtest_gallery} we compare images obtained by modeling discrete sources using inner \uv\ cuts of 1500, 2500, 5000, 8000, and 10000$\lambda$ (from \textit{left} to \textit{right}). As evident, the residual contribution from partially subtracted discrete sources increases when progressively larger inner \uv\ cuts are adopted. Images produced with smaller inner \uv\ cuts instead appear cleaner, as emission on larger scales is included in the model and subsequently subtracted. Nonetheless, manually excluding the extended emission associated with the ICM from these models before prediction is challenging, as it can blend with the emission from discrete sources. As a result, part of the cluster emission may still be included in the model and subtracted, leading to an artificial smoothing of the diffuse cluster emission. In this paper we adopted as reference the subtraction obtained by modeling the discrete sources with an inner \uv\ cut of 5000$\lambda$. Residual contamination from partially subtracted resolved sources was subsequently masked in the image plane prior to the SIG analysis. \\
\indent
Once discrete sources are subtracted in the visibility plane, low-resolution imaging is typically performed to enhance the detection of diffuse emission. Also in this step, the choice of the inner \uv\ cut adopted during imaging affects the recovered extended emission, as it determines the ability of the interferometer to recover structures on large angular scales. In this case, the adopted value represents a trade-off between recovering as much of the extended cluster emission as possible and filtering out large-scale structures arising from the shortest baselines, where calibration is more challenging and contamination from Galactic foreground emission can become significant \citep[see Appendix~A in][]{botteon25}. This effect is also observed in our images of A2255 shown in Fig.~\ref{fig:shortbaselinestest_gallery}, where inner \uv\ cuts of 20, 30, 40, 50, 60, and 70$\lambda$ (from \textit{top left} to \textit{bottom right}) were adopted during imaging. The large-scale pattern (characterized by emission ``patches'' and/or ``stripes'') detected in these images is likely of Galactic origin, as seen in many \lotss\ pointings when low-resolution imaging with discrete sources subtracted is performed (Oei et al., in prep.). With shorter inner \uv\ cuts, the region between the NE and NW relics appears more filled with diffuse emission, suggesting the presence of genuine cluster emission. However, this region is also crossed by a large-scale stripe unrelated to the cluster; therefore, the images presented in this paper were obtained adopting an inner \uv\ cut of 60$\lambda$ as the reference value.

\subsection{SIG}\label{app:sig}

\begin{figure*}
  \centering
  \includegraphics[width=.24\hsize,trim={0cm 0cm 0cm 0cm},clip,valign=c]{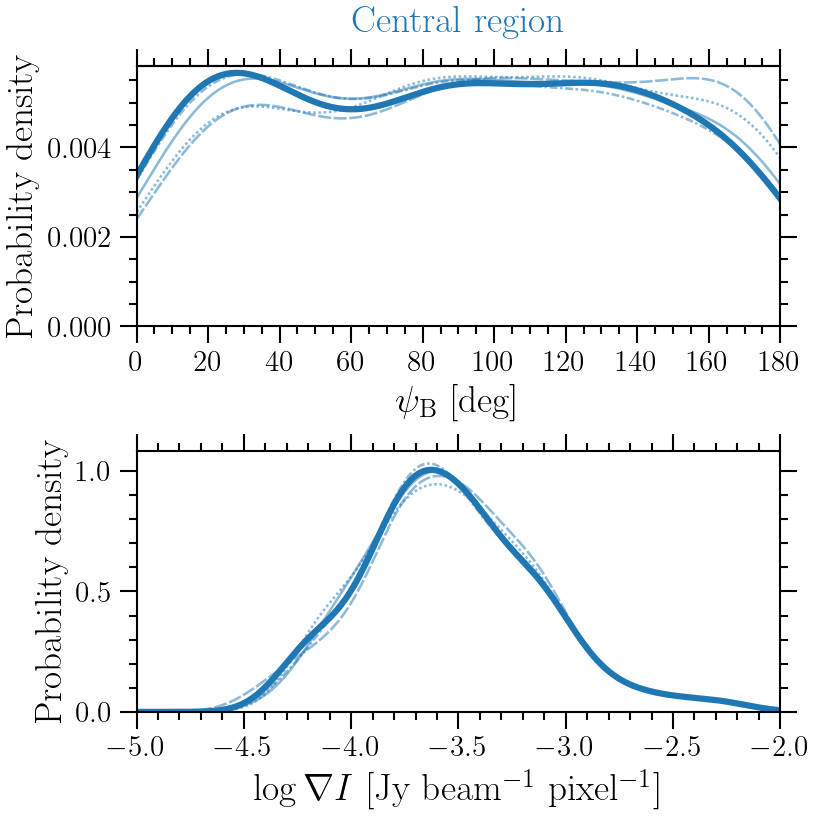}
  \includegraphics[width=.24\hsize,trim={0cm 0cm 0cm 0cm},clip,valign=c]{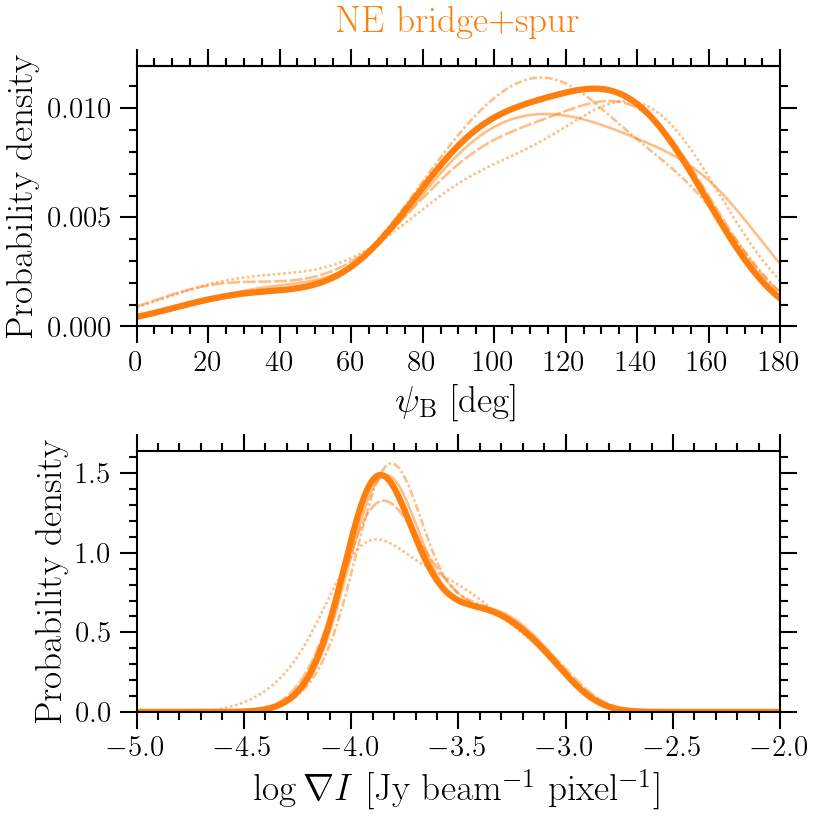}
  \includegraphics[width=.24\hsize,trim={0cm 0cm 0cm 0cm},clip,valign=c]{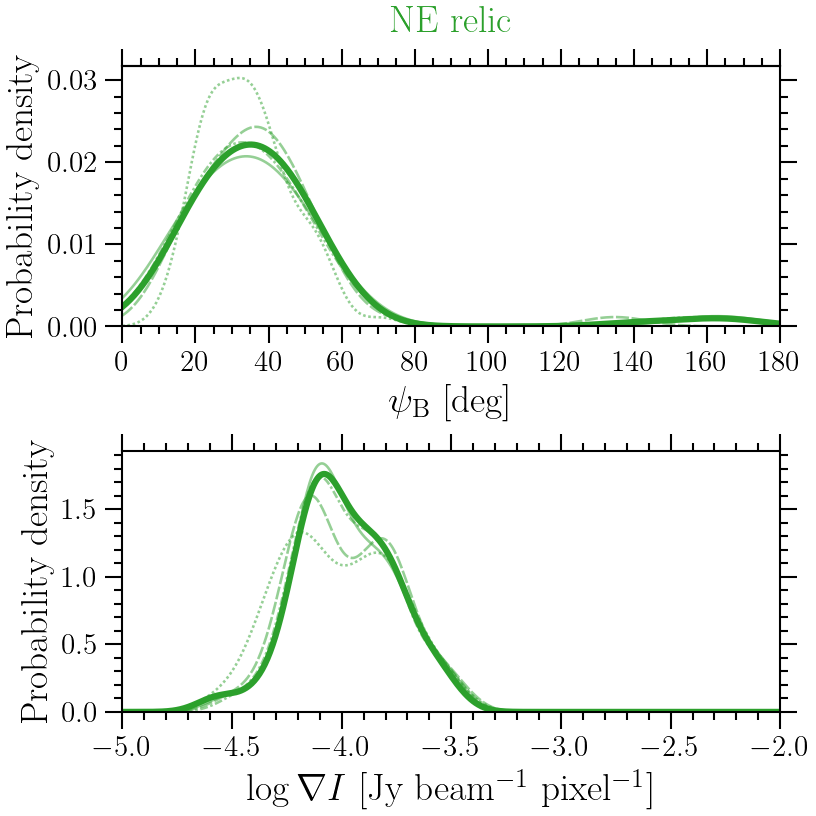}
  \includegraphics[width=.24\hsize,trim={0cm 0cm 0cm 0cm},clip,valign=c]{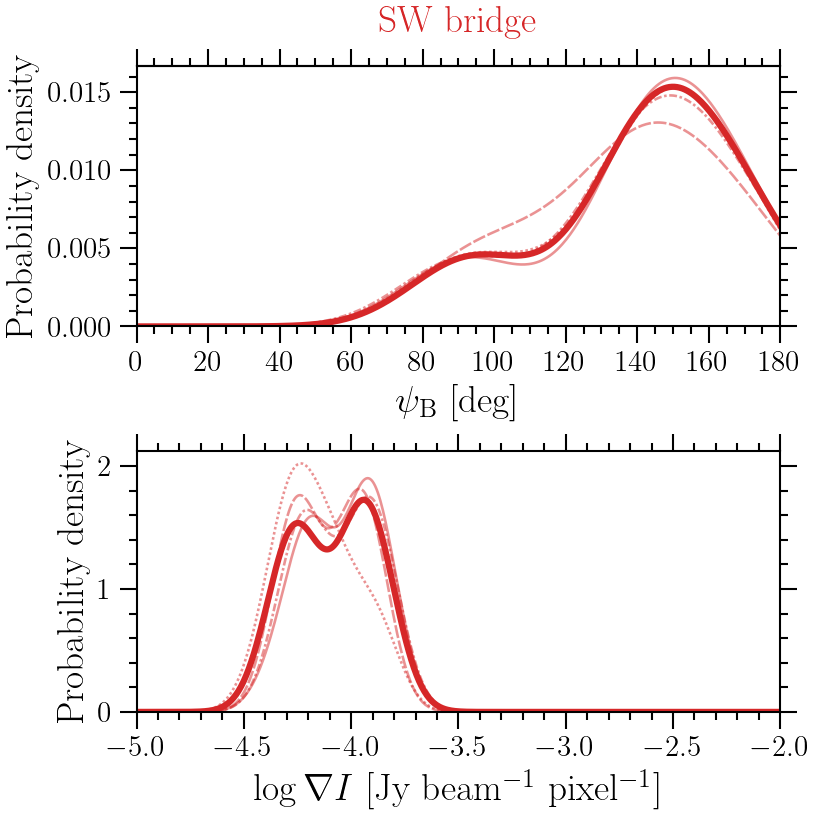} \\ \vspace{1ex}
  \includegraphics[width=.24\hsize,trim={0cm 0cm 0cm 0cm},clip,valign=c]{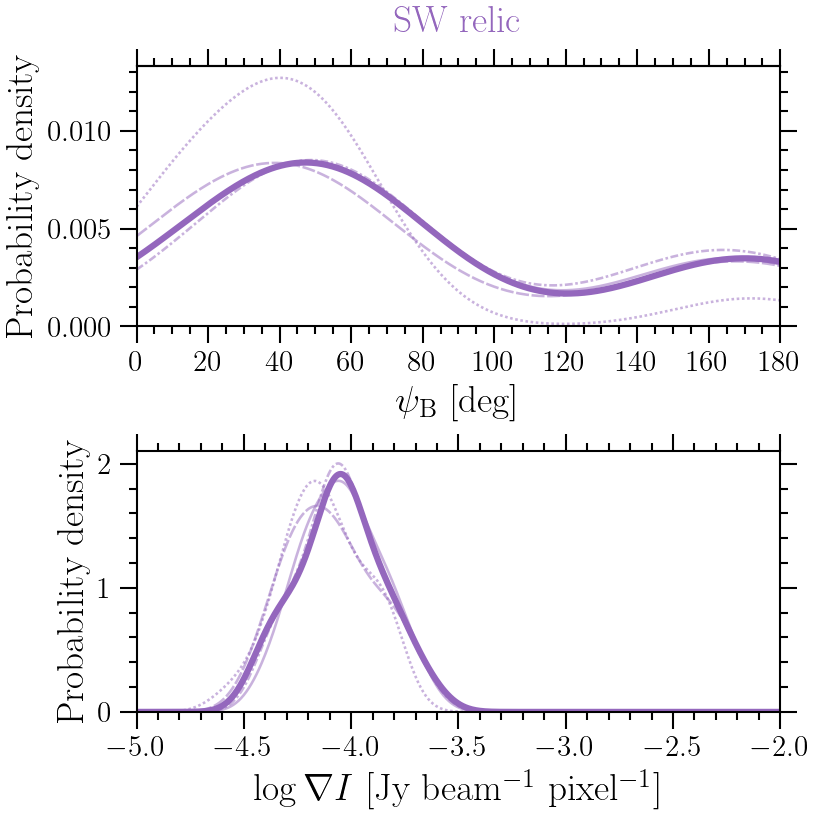}
  \includegraphics[width=.24\hsize,trim={0cm 0cm 0cm 0cm},clip,valign=c]{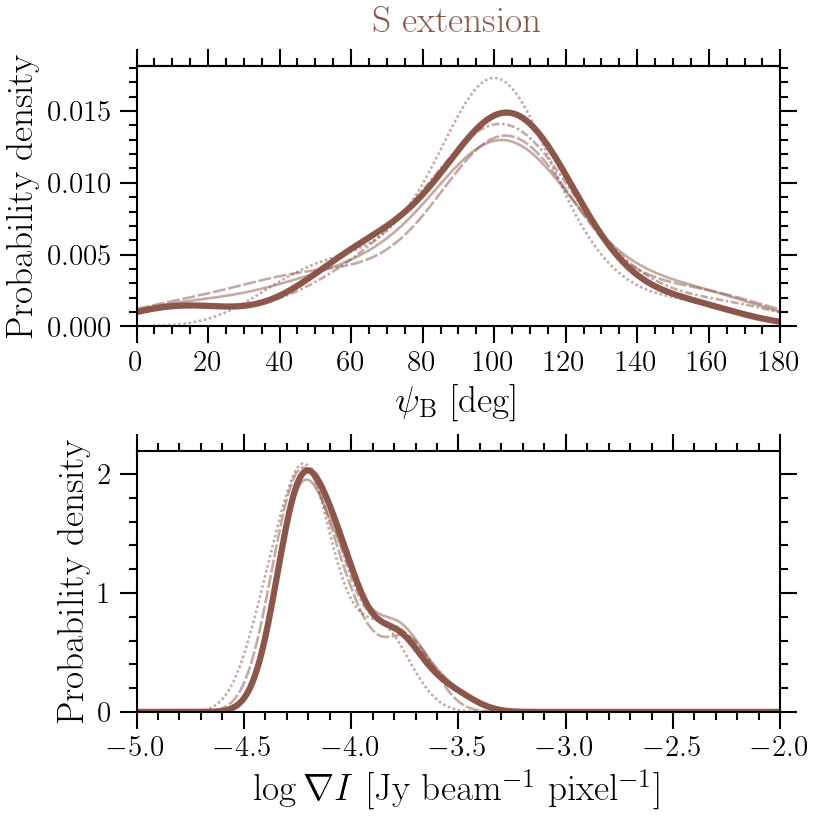}
  \includegraphics[width=.24\hsize,trim={0cm 0cm 0cm 0cm},clip,valign=c]{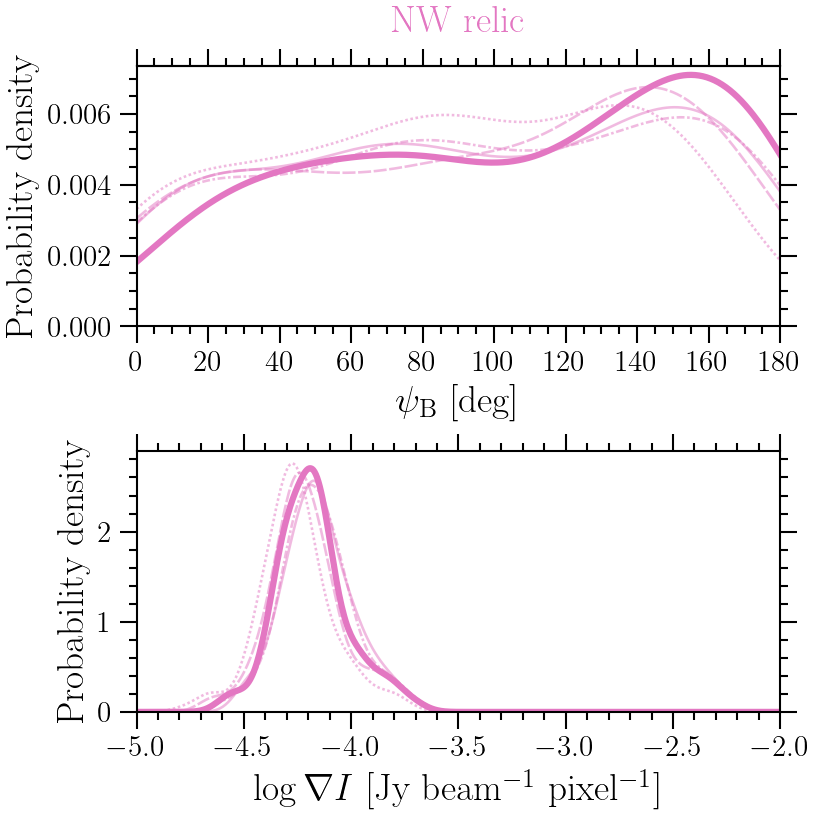}
  \includegraphics[width=.24\hsize,trim={0cm 0cm 0cm 0cm},clip,valign=c]{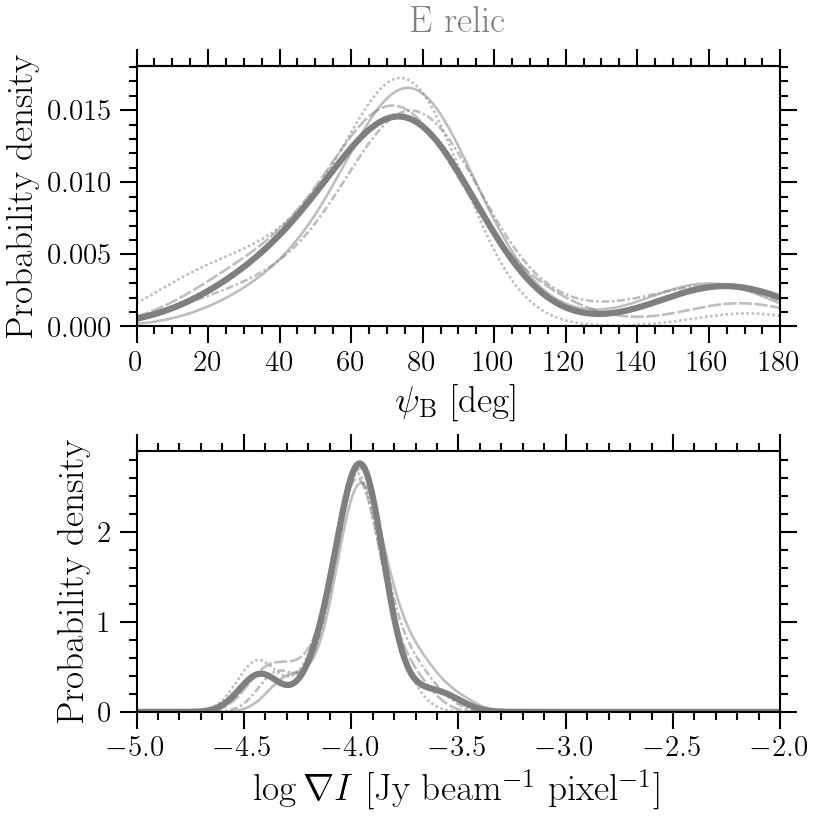}
  \caption{Impact of the choice of the inner \uv\ cut for discrete source subtraction on the SIG results. Distributions of gradient angles and amplitudes are shown as kernel density estimation curves. The thick solid line represents the result from the reference image, where discrete sources were subtracted using an inner \uv\ cut of 5000$\lambda$.}
  \label{fig:subtest_SIG}
\end{figure*}

\begin{figure*}
  \centering
  \includegraphics[width=.24\hsize,trim={0cm 0cm 0cm 0cm},clip,valign=c]{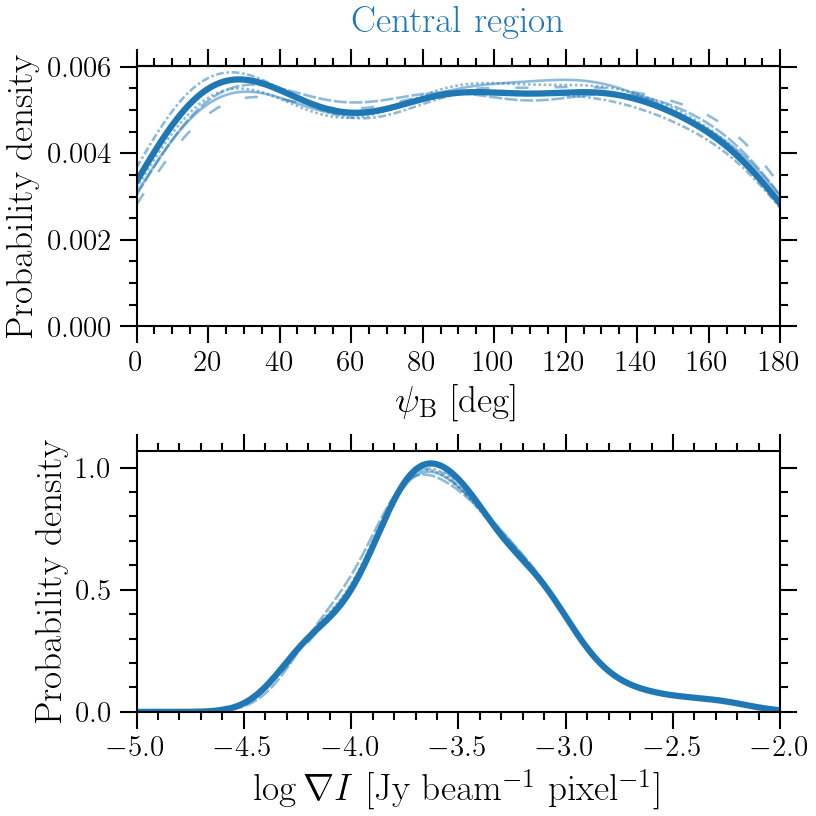}
  \includegraphics[width=.24\hsize,trim={0cm 0cm 0cm 0cm},clip,valign=c]{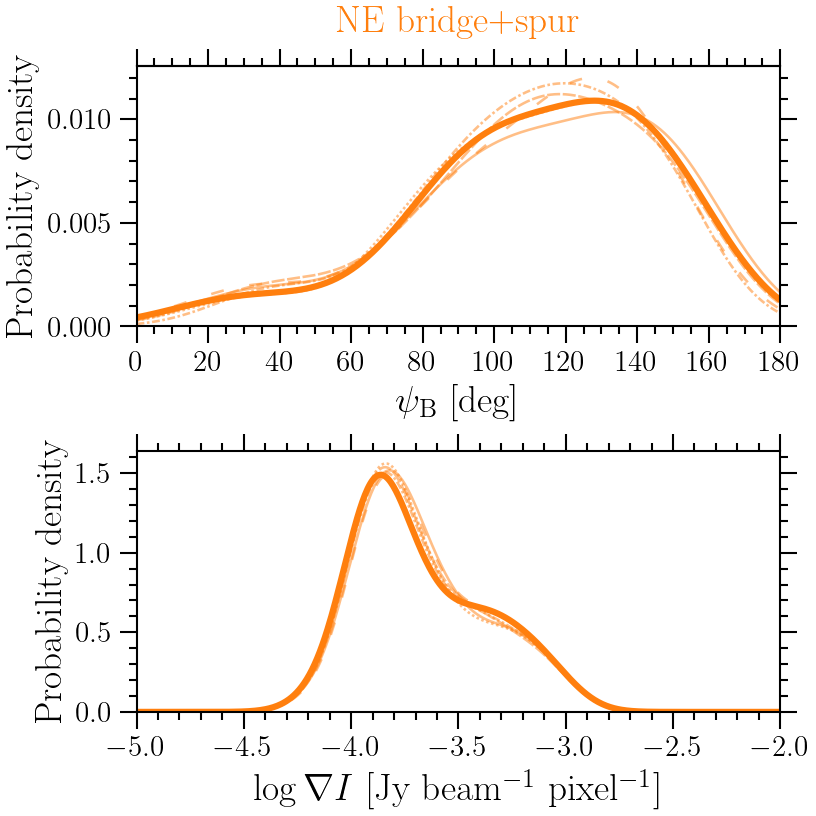}
  \includegraphics[width=.24\hsize,trim={0cm 0cm 0cm 0cm},clip,valign=c]{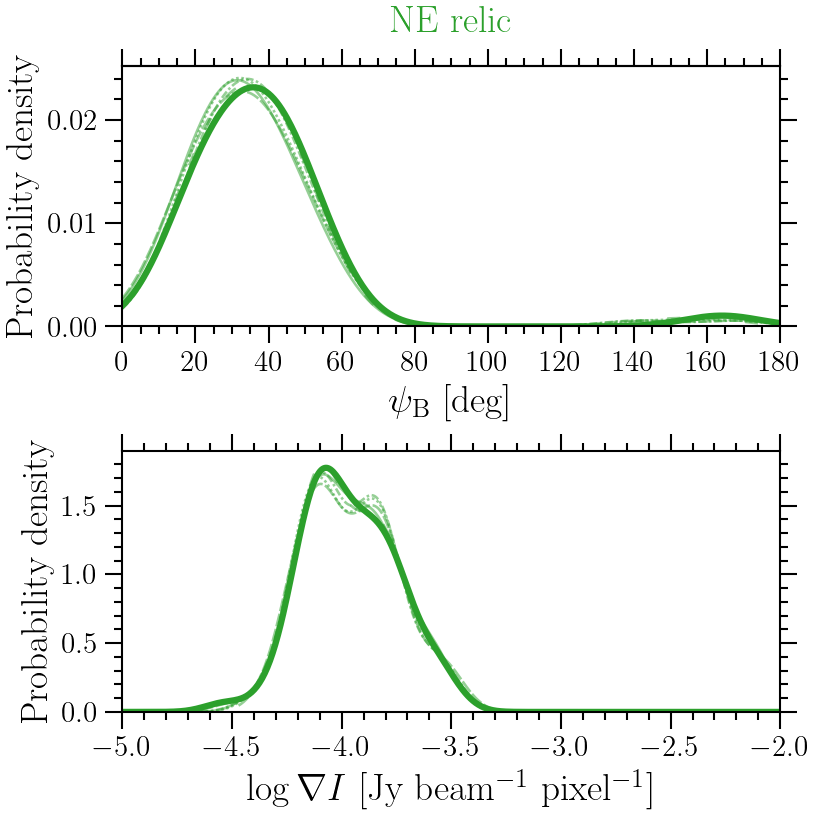}
  \includegraphics[width=.24\hsize,trim={0cm 0cm 0cm 0cm},clip,valign=c]{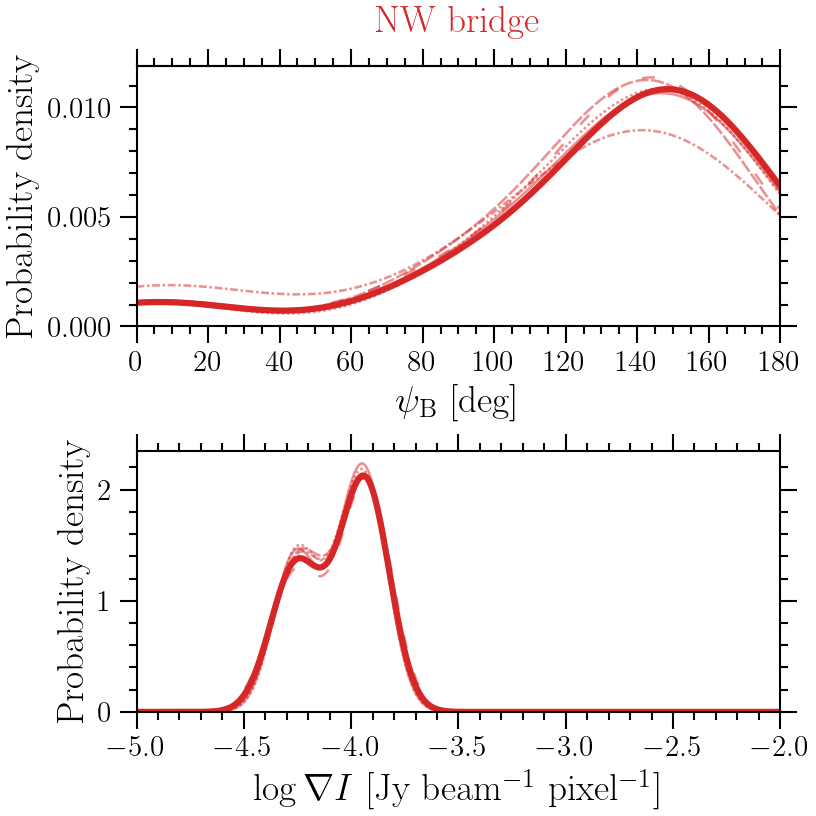} \\ \vspace{1ex}
  \includegraphics[width=.24\hsize,trim={0cm 0cm 0cm 0cm},clip,valign=c]{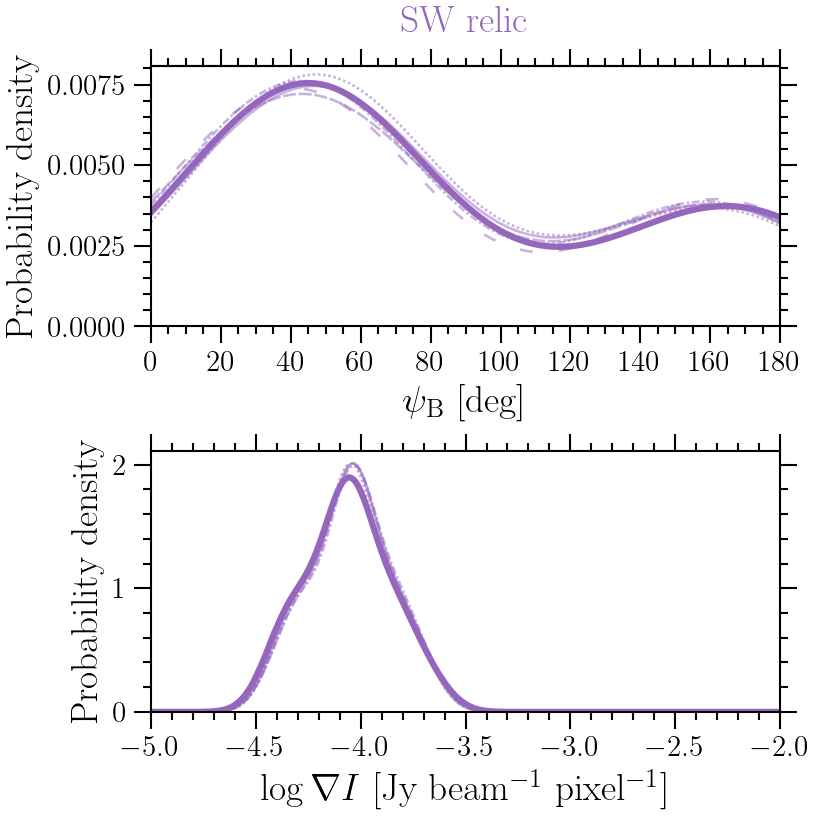}
  \includegraphics[width=.24\hsize,trim={0cm 0cm 0cm 0cm},clip,valign=c]{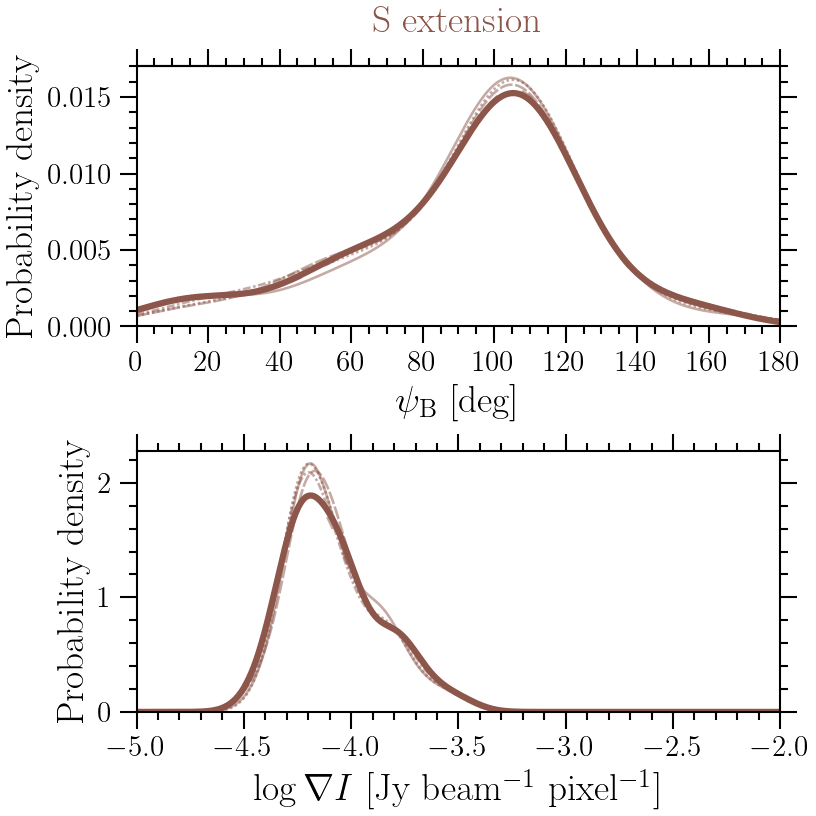}
  \includegraphics[width=.24\hsize,trim={0cm 0cm 0cm 0cm},clip,valign=c]{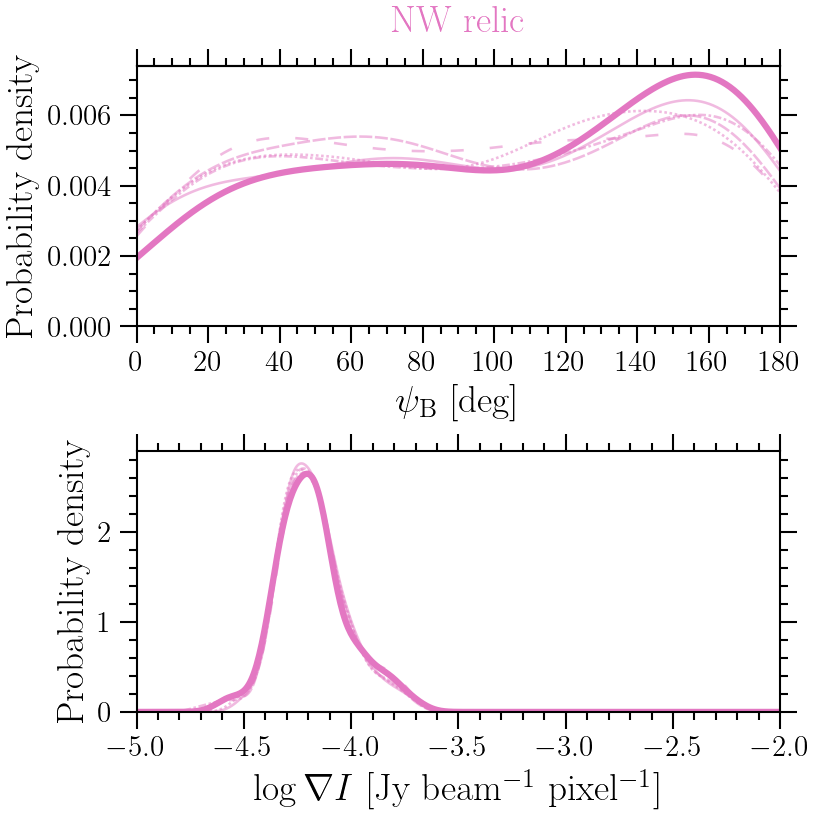}
  \includegraphics[width=.24\hsize,trim={0cm 0cm 0cm 0cm},clip,valign=c]{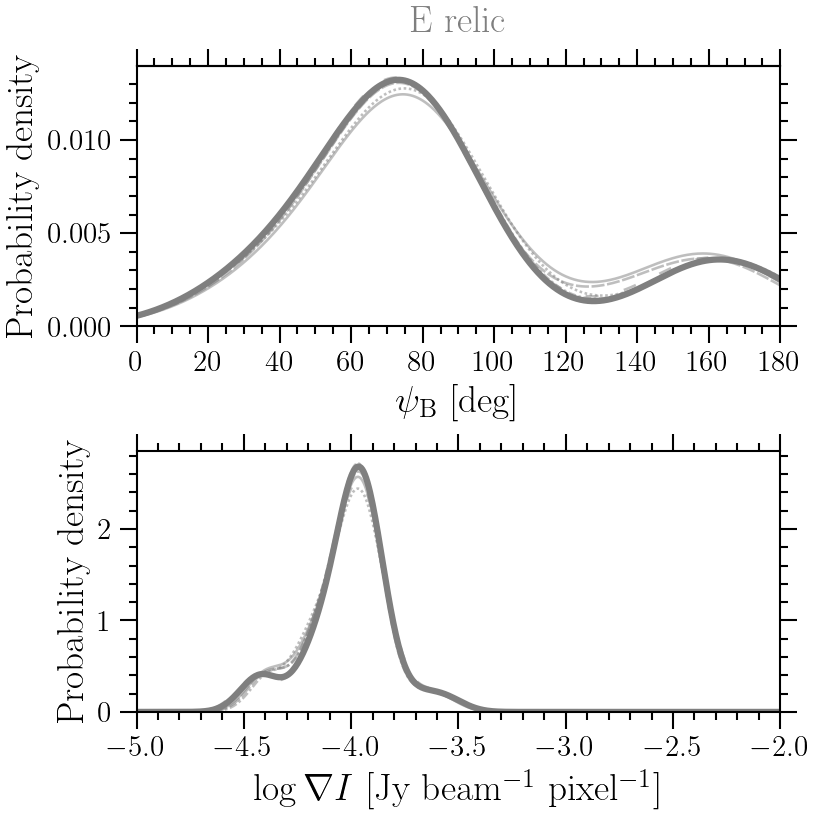}
  \caption{Impact of the choice of the inner \uv\ cut for the recovery of extended emission on the SIG results. Distributions of gradient angles and amplitudes are shown as kernel density estimation curves. The thick solid line represents the result from the reference image, obtained by adopting an inner \uv\ cut of 60$\lambda$ during imaging.}
  \label{fig:shortbaselinestest_SIG}
\end{figure*}

The motivation for producing the sets of images discussed in the previous section is to assess whether different imaging settings affect the results obtained from the SIG analysis. We therefore repeated the analysis presented in Section~\ref{sec:sig_a2255} on 35\arcsec-resolution images with discrete sources subtracted, adopting different \uv\ cuts for the modeling of discrete sources and for the recovery of the extended emission. In Figs.~\ref{fig:subtest_SIG} and \ref{fig:shortbaselinestest_SIG} we report the distributions of gradient angles and amplitudes within the eight regions analyzed in the main text. A common mask including only pixels where SIG results are available in all images was applied to ensure that the distributions are directly comparable. To facilitate comparison, the distributions were smoothed using a Gaussian kernel density estimator, with the solid thick line representing the reference image. The global trends of the distributions are not strongly affected by the different imaging parameters, with orientation angles being more sensitive than amplitudes to these choices. \\
\indent
For a more quantitative comparison, we computed the relative angle between the magnetic field orientation obtained from the additional images, $\psi_{\rm B}(x,y)$, and that obtained from our reference image, $\psi^{\prime}_{\rm B}(x,y)$, as

\begin{equation}\label{eq:theta_r}
 \theta_{\rm r}(x, y) = \psi_{\rm B}(x,y) - \psi^{\prime}_{\rm B}(x,y),
\end{equation}

\noindent
and defined the alignment measure (AM) between the two gradients as

\begin{equation}\label{eq:am}
 {\rm AM}(x,y) = 2[\cos^2 \theta_{\rm r}(x,y) - 1/2],
\end{equation}

\noindent
which quantifies the alignment of two SIG results. The same definition has been adopted to compare the correspondence between velocity gradients and the magnetic field \citep{gonzalezcasanova17} or the magnetic field orientation derived from the SIG analysis and that obtained from polarization \citep{hu24clusters}. As defined in Eq.~\ref{eq:am}, ${\rm AM}=1$ corresponds to perfect parallel alignment, while ${\rm AM}=-1$ indicates perpendicular alignment. We therefore computed the AM of the magnetic field orientations derived for the eight regions analyzed in A2255 using different imaging settings, and summarized the results in Tab.~\ref{tab:AM}, where we report the median values of the corresponding AM distributions. As ${\rm AM} \approx 1$, we conclude that the SIG results reported in the paper do not significantly depend on the inner \uv\ cuts adopted for the modeling of discrete sources and the recovery of the extended emission.

\begin{table}[h]
\centering
\caption{Median values of AM relative to the reference image obtained by modeling discrete sources with an inner \uv\ cut of 5000$\lambda$ and imaging the residual data with an inner \uv\ cut of 60$\lambda$.}
\label{tab:AM}
\begin{tabular}{lcccc|ccccc}
\hline
\hline
 Region & \multicolumn{4}{c}{Subtraction test} & \multicolumn{5}{c}{Extended emission test} \\
       & 1500$\lambda$ & 2500$\lambda$ & 8000$\lambda$ & 10000$\lambda$ & 20$\lambda$ & 30$\lambda$ & 40$\lambda$ & 50$\lambda$ & 70$\lambda$ \\
\hline
  Halo & 1.00 & 1.00 & 1.00& 1.00 & 1.00 & 1.00 & 1.00& 1.00 & 1.00\\
  NE bridge+spur & 0.99 & 1.00 & 1.00 & 1.00 & 1.00 & 1.00 & 1.00& 1.00 & 1.00\\
  NE relic & 1.00 & 1.00 & 1.00 & 1.00 & 1.00 & 1.00 & 1.00& 1.00 & 1.00\\
  SW bridge & 1.00 & 0.99 & 1.00 & 1.00 & 1.00 & 1.00 & 1.00& 1.00 & 1.00\\
  SW relic & 0.95 & 0.99 & 1.00 & 1.00 & 1.00 & 0.99 & 0.99 & 0.99 & 0.99\\
  S extension & 0.99 & 0.99 & 1.00 & 1.00 & 1.00 & 1.00 & 1.00& 1.00 & 1.00\\
  NW relic & 0.96 & 0.99 & 1.00 & 0.99 & 1.00 & 0.99 & 0.99 & 1.00 & 0.99\\
  E relic & 0.96 & 0.96 & 1.00 & 1.00  & 1.00 & 1.00 & 1.00& 1.00 & 1.00\\
\hline
\end{tabular}
\end{table}

\newpage
\section{Supplementary images of the candidate E relic}\label{app:E_relic}

\begin{figure*}
  \centering
  \includegraphics[width=\hsize,trim={0cm 0cm 0cm 0cm},clip,valign=c]{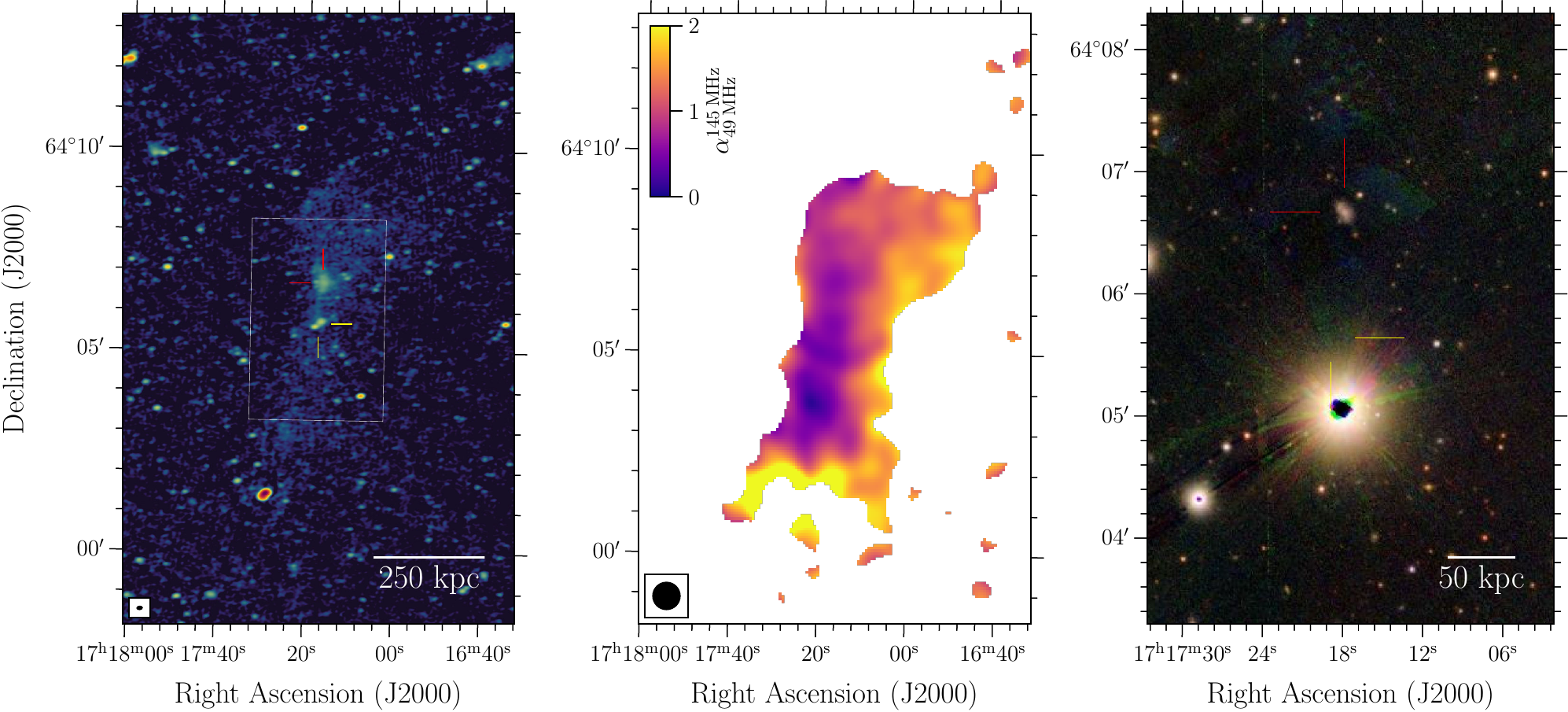}
  \caption{The candidate E relic in A2255. \textit{Left panel}: zoom-in of the Ultra-Deep Field full-resolution image shown in Fig.~\ref{fig:robust-0.5}. \textit{Central panel}: spectral index ($S_\nu \propto \nu^{-\alpha}$ convention) map between 49 and 145 MHz at 40\arcsec\ resolution, with discrete sources subtracted, from \citet{botteon22a2255}. \textit{Right panel}: optical image of the FoV outlined by the rectangular box in the left panel, from \panstarrs\ \textit{gri} \citep{chambers16arx}. Red and yellow segments mark the positions of compact radio components likely associated with a spiral galaxy (with a photometric-$z$ of $0.148 \pm 0.040$ from \citealt{duncan22}) and a double-lobed radio galaxy, respectively; the optical counterpart of the latter is obscured by a bright foreground star. The scale bars indicate the physical scale at the redshift of A2255.}
  \label{fig:E_relic}
\end{figure*}

\end{appendix}

\end{document}

%% file: latex_mycommands.txt
\def \eg {e.g.}

\def \ie {i.e.}
\def \cf {cf.}

\def \omegam {{\hbox{$\Omega_{\rm m}$}}}
\def \omegal {{\hbox{$\Omega_\Lambda$}}}
\def \hzero {{\hbox{$H_0$}}}
\def \arcmin {\hbox{$^\prime$}}
\def \arcsec {\hbox{$^{\prime\prime}$}}
\def \deg {\hbox{$^\circ$}}

\def \mach {{\hbox{$\mathcal{M}$}}}

\def \rfive {\hbox{$r_{500}$}}

\def \rtwo {\hbox{$r_{200}$}}
\newcommand{\alfven }{Alfv\'{e}n}
\newcommand{\alfvenic }{Alfv\'{e}nic}

\newcommand{\kmsmpc }{\mbox{km s$^{-1}$ Mpc$^{-1}$}}

\newcommand{\mujyb }{\mbox{$\mu$Jy beam$^{-1}$}}

\newcommand{\muG }{\mbox{$\mu$G}}

\newcommand{\uv }{\textit{uv}}

\newcommand{\wsclean }{\textsc{WSClean}}

\newcommand{\aoflagger }{\textsc{AOFlagger}}
\newcommand{\dysco }{\textsc{Dysco}}

\newcommand{\dpppE }{Default PreProcessing Pipeline}

\newcommand{\xrism }{{\em XRISM}}

\newcommand{\vla }{VLA}

\newcommand{\lofar }{LOFAR}

\newcommand{\ska }{SKA}
\newcommand{\skaE }{Square Kilometer Array}

\newcommand{\askap }{ASKAP}

\newcommand{\meerkat }{MeerKAT}
\newcommand{\wsrt }{WSRT}

\newcommand{\lotss }{LoTSS}
\newcommand{\lotssE }{LOFAR Two-meter Sky Survey}

\newcommand{\tgss }{TGSS}
\newcommand{\tgssE }{TIFR GMRT Sky Survey}

\newcommand{\panstarrs }{Pan-STARRS}